\documentclass[AMA,STIX2COL]{MRM}
\articletype{Research Article}%
\input{shortcuts}
\received{NA}
\revised{NA}
\accepted{NA}
\usepackage{mathtools}
\usepackage{needspace}
\usepackage{amsmath}
\usepackage{adjustbox}        
\usepackage{caption}
\usepackage{subcaption}
\usepackage{enumitem}
\usepackage{pdfcomment}

\usepackage[capitalise,nameinlink]{cleveref}
\usepackage{marginnote}

\reversemarginpar

\usepackage[dvipsnames]{xcolor}

\newcounter{algo}
\renewcommand{\thealgo}{\arabic{algo}}
\newenvironment{myalgorithm}[1]{%
    \refstepcounter{algo}%
    \hrule%
    \vspace{0.5mm}%
    \paragraph{Algorithm \thealgo.\ #1}%
    \vspace{0.5mm}%
    \hrule%
    \vspace{1mm}%
}{%
    \vspace{1mm}%
    \hrule%
}
\algnewcommand{\Constant}{\item[\textbf{Constants:}]}

\renewcommand{\thefootnote}{\fnsymbol{footnote}}

\makeatletter
\newcommand\blfootnote[1]{%
  \begingroup
    \renewcommand\thefootnote{}\footnote{#1}%
    \addtocounter{footnote}{-1}%
  \endgroup
}
\makeatother

\begin{document}

\title{Accelerating MRI with Longitudinally-informed Latent Posterior Sampling}

\author[1]{Yonatan Urman$^{*,}$}{\orcid{0000-0002-5763-8174}}

\author[1]{Zachary Shah$^{*,}$}{\orcid{0009-0006-8652-1924}}

\author[2]{Ashwin Kumar}{\orcid{0009-0002-9859-4275}}

\author[2]{Bruno P. Soares}{\orcid{0000-0003-4185-6940}}

\author[1,2]{Kawin Setsompop}{\orcid{0000-0003-0455-7634}}

\authormark{URMAN, SHAH \textsc{et al}}

\address[1]{\orgdiv{Electrical Engineering}, \orgname{Stanford University}, \orgaddress{\state{California}, \country{USA}}}

\address[2]{\orgdiv{Radiology}, \orgname{Stanford University}, \orgaddress{\state{California}, \country{USA}}}

\corres{Zachary Shah: \email{zshah9@stanford.edu}}

\abstract[Abstract]{
\section{Purpose}
To accelerate MRI acquisition by incorporating the previous scans of a subject during reconstruction. Although longitudinal imaging constitutes much of clinical MRI, leveraging previous scans is challenging due to the complex relationship between scan sessions, potentially involving substantial anatomical or pathological changes, and the lack of open-access datasets with both longitudinal pairs and raw k-space needed for training deep learning-based reconstruction models.
\section{Methods}
We propose a diffusion-model-based reconstruction framework that eliminates the need for longitudinally paired training data. During training, we treat all scan timepoints as samples from the same distribution, therefore requiring only standalone images. At inference, our framework integrates a subject's prior scan in magnitude DICOM format, which is readily available in clinical workflows, to guide reconstruction of the follow-up. To support future development, we introduce an open-access clinical dataset containing multi-session pairs including prior DICOMs and follow-up k-space.
\section{Results}
Our method consistently outperforms both longitudinal and non-longitudinal baseline reconstruction methods across various accelerated Cartesian acquisition strategies. In imaging regions highly similar to the prior scan, we observe up to 10\% higher SSIM and 2 dB higher PSNR, without degradation in dissimilar areas. Compared to longitudinal reconstruction baselines, our method demonstrates robustness to varying degrees of anatomical change and misregistration.
\section{Conclusion}
We demonstrate that prior scans can be effectively integrated with state-of-the-art diffusion-based reconstruction methods to improve image quality and enable greater scan acceleration, without requiring an extensive longitudinally-paired training dataset.
\vspace{-4em}
}


\maketitle
\blfootnote{\textsuperscript{*}These authors contributed equally to this work.}

\clearpage

\jnlcitation{\cname{%
\author{Urman Y}, 
\author{Shah Z}, 
\author{Kumar A}, 
\author{Soares B.P}, and 
\author{Setsompop K}} (\cyear{2025}), 
\ctitle{Accelerating MRI with Longitudinally-informed Latent Posterior Sampling}, \cjournal{Magn. Reson. Med.}.}

\vspace{-1em}\section{Introduction}\label{sec:intro}
Scan durations in MRI have decreased significantly in the past decade, notably due to advances in compressed sensing (CS)\cite{lustig2007sparse} and deep learning (DL)-based reconstruction methods\cite{heckel2024deep,fastmri}. Despite these improvements, MRI still remains a relatively slow imaging modality\cite{van2019value}, limiting patient comfort and clinical throughput. At the same time, its clinical relevance continues to grow\cite{bor2021increasing}.

Compared to early hand-crafted priors that exploit image sparsity\cite{joshi2009mri,lustig2007sparse}, DL-based priors\cite{aggarwal2018modl,sriram2020,jalal2021robust} learn data distributions, forming general regularizers over MR image features. Such methods are now approaching the achievable limit of scan acceleration attainable through learned regularizers alone, motivating the use of additional patient-specific information in reconstruction. One well-established example is contrast-to-contrast translation, where the reconstruction problem is conditioned on a different contrast from the same session. Leading DL approaches for this task leverage supervised frameworks trained on multi-contrast k-space datasets\cite{levac2023mri, atalık2025trust}, of which several are publicly available \cite{dsai2021skm, lyu2023m4raw}. 

An underexplored regularizer of comparable potential is previous scans of the same subject. MRI is often used in longitudinal studies \cite{Rees2009,zahid2022impact,Weizman2014,Young2011}, where patients are imaged repeatedly over time. These prior scans are typically highly correlated with newly acquired images but are rarely incorporated into the reconstruction or acquisition process of follow-up studies.

At first glance, longitudinal reconstruction bears similarity to contrast-to-contrast translation, where the goal is to reconstruct one scan given another. Longitudinal reconstruction nevertheless faces a different set of practical challenges. The conventional DL approach would be to supervise a regularizer over all possible longitudinal changes, requiring datasets of scan timepoint pairs with raw multi-channel k-space data for end-to-end training. However, to our knowledge, existing public longitudinal datasets \cite{lamontagne2019oasis, hawco2022longitudinal, suter2022lumiere, petersen2010alzheimer} only include magnitude images, typically in DICOM format, which are appropriate for post-processing applications but insufficient for training complex-valued reconstruction models\cite{shimron2022datacrimes}. Even if complex-valued longitudinal datasets were readily available, accurately modeling all longitudinal anatomical and pathological evolutions would mandate an order of magnitude increase to the training data requirement, challenging the practicality and generalizability of conditionally supervised approaches for this problem. 

With this challenge in mind, several approaches have proposed relying on prior scans only at inference. LACS~\cite{weizman2015compressed} adds an adaptive regularization enforcing sparse differences to the prior scan, improving the standard CS reconstruction. A different approach, NERP\cite{shen2022nerp}, leverages the prior scan to better condition an implicit neural representation (INR) based reconstruction, showing improvement in CT with a preliminary extension to MRI. Ultimately, such approaches still underperform state-of-the-art supervised DL methods agnostic of any conditioning scans. Additionally, their fidelity is sensitive to registration and structural change between scan pairs, as both methods exploit pixel-wise similarity in their respective regularization modalities.

Hence, we observe that modeling general longitudinal regularizers is challenging both theoretically (an involute conditional distribution) and practically (inaccessible data requirement). Though longitudinal imaging is a conditional problem, we propose that, when viewed independently, the prior $\xp$ and target $\bx$ reconstructions can be modeled as samples from the same marginal distribution: $\xp \sim p(\bx), \bx \sim p(\bx)$. This symmetry allows us to avoid explicitly modeling the conditional $p(\bx \mid \xp)$, utilizing unconditional approaches to instead supervise $p(\bx)$ directly. With this insight, we formulate a longitudinal reconstruction framework that does not require longitudinally paired training data, but can exploit a deep general prior in conjunction with a previous scan in magnitude DICOM format observed at inference alone.

To accomplish this, we use a diffusion model prior, which has state-of-the-art capability to represent an image distribution as a deep regularizer independently of the forward imaging model. We train the model on unpaired data, consisting of standalone images rather than longitudinal pairs. We specifically employ a latent diffusion model (LDM) to redirect regularization from pixel-space to latent-space. This enables more efficient use of prior scan features and improves robustness to misregistration and morphological changes.

To leverage the prior scan, we introduce a hot-start initialization strategy, where reconstruction begins from a noisy version of the prior scan’s latent representation. The noise level is set based on scan similarity estimated from a fast preliminary reconstruction, allowing the method to balance reliance on the prior with flexibility for new or altered content. This keeps reconstructions consistent with prior information while avoiding the need to explicitly model the conditional distribution.

In summary, the main contributions of this work are:
\begin{enumerate}[leftmargin=*]
\item We propose a forward-model-agnostic DL framework for longitudinal MRI reconstruction that eliminates the need for paired training data. We show that incorporating the prior boosts qualitative and quantitative performance, extending the achievable neuroimaging acceleration rate for 1D variable-density undersampling from 5–7$\times$ to 9$\times$, and for 2D variable-density undersampling from 20-23$\times$ to 30$\times$, while maintaining reconstruction quality comparable to unconditional baselines at lower accelerations.
\item Compared to longitudinal reconstruction baselines, our method shows increased robustness to misregistration and varying degrees of change between scans, with a single trained model generalizing to diverse clinical applications.
\item To support future research, we curated and are actively expanding a multi-contrast multi-session brain MRI dataset, which currently contains 232 scans from 53 subjects. The dataset includes raw k-space data with corresponding prior images, which we have publicly released\footnote{Data available at https://doi.org/10.25740/rq296rb2765.} for further longitudinal reconstruction benchmarking.
\end{enumerate}

\noindent A preliminary version of this work appears in ISMRM 2025\cite{shah2025pips}.

\vspace{-1em}\section{Theory}\label{sec:theory}
This section begins with a brief background on LDM-based reconstruction, on which the current work is based. We then motivate and develop a framework for hot-starting diffusion model sampling from a given image within the model distribution, which will later be used to introduce our proposed method.

\vspace{-1em}\subsection{Background}\label{sec:background}
MRI reconstruction is commonly formulated as a maximum a posteriori problem:
\begin{equation}\label{eq:map}
\xgt = \argmax{\bx}{\;\log p(\by \mid \bx) + \log p(\bx)}.
\end{equation}
Here, $\by = \bcA(\bx) + \bn$ are undersampled, noisy measurements, with $\bcA$ denoting the MRI forward model and $\bn$ is Gaussian noise. The first term in Eq.~\eqref{eq:map} leads to a data consistency (DC) loss $\|\bcA(\bx) - \by\|_2^2$, while the second term acts as a prior to regularize ill-posed reconstructions.

\vspace{-0.75em}\subsubsection{Score-based Diffusion Models}\label{sec:score_models}
To learn a deep regularizer for Eq.~\eqref{eq:map}, we need to model the image prior $p(\bx)$. Score-based diffusion models\cite{ho2020denoising, song2021scorebased} have recently emerged as state-of-the-art for this task. They define a forward process that gradually perturbs data into Gaussian noise, and learn to reverse it. Generation is performed by solving the reverse-time stochastic differential equation:
\begin{equation}\label{eq:reverse_diff}
d\bx_t = \left[f(t)\bx_t - g^2(t)\nabla_{\bx_t}\log p(\bx_t)\right] dt + g(t)d\bw_t
\end{equation}
where $f(t)$ and $g(t)$ are time-dependent drift and diffusion coefficients, $\bw_t$ is a Wiener process, and the score function $\nabla_{\bx} \log p_t(\bx)$ is approximated by a neural network $\bs_\theta^\star(\bx_t, t)$ trained via denoising score matching\cite{vincent2011score}. Notably, training is agnostic to $\bcA$; i.e., a single learned model can be applied broadly across various reconstruction tasks.

In DDPM\cite{ho2020denoising} the reverse process is typically discretized into $T = 1000$ steps, forming noisy intermediates $\{\bx_t\}_{t=0}^{T}$, where $\bx_0 \sim p(\bx)$. Variants like DDIM\cite{song2020denoising} accelerate high-quality sampling to as few as 50 steps.

\vspace{-0.75em}\subsubsection{Latent Diffusion Models}\label{sec:ldms}
Modeling $p(\bx)$ in high-dimensional image space (i.e., total number of voxels) requires learning feature-based data representations alongside the underlying distribution, which is intensive in both data and compute. LDMs\cite{rombach2022highresolution} address this by first training a Variational Autoencoder (VAE)\cite{kingma2013auto} to map images into a compact latent space $p(\bz)$ using an encoder–decoder pair $(\cE, \cD)$. For 2D complex MRI images of shape $N \times M$:
\begin{align}\label{eq:vae}
\bitsmall
&\cE: \bx \in \C^{N\times M} \rightarrow \bz \in \R^{\frac{N}{K}\times \frac{M}{K} \times C}\\
&\cD: \bz \in \R^{\frac{N}{K}\times \frac{M}{K} \times C} \rightarrow \bx \in \C^{N\times M}
\normalsize
\end{align}
where $K$ is a spatial down-sampling factor. A diffusion model is then trained in this latent space, and samples are decoded back as images via $\bx_0 = \cD(\bz_0)$.
This setup enables learning in a lower-dimensional, semantically meaningful space, accelerating training and improving image synthesis.

\vspace{-0.75em}\subsubsection{Posterior Sampling: Solving Inverse Problems with Diffusion Models}\label{sec:posterior_sampling}
To reconstruct images given a diffusion model and $\by$, we use posterior sampling~\cite{song2021solving, jalal2021robust}, which aims to draw samples from $p(\bx \mid \by)$. Since diffusion sampling relies on the score function (Eq.~\eqref{eq:reverse_diff}), Bayes’ rule gives:
\begin{equation}
\label{eq:posterior}
\nabla_{\bx_t} \log p(\bx_t \mid \by) = 
    \underbrace{\nabla_{\bx_t} \log p(\bx_t)}_\text{Prior Score}
    + 
    \underbrace{\nabla_{\bx_t} \log p(\by \mid \bx_t)}_\text{Likelihood Score}
\end{equation}
The prior score is provided by a pre-trained diffusion model, and the likelihood score enforces consistency with the measurements.

As $\by$ depends on the denoised $\bx_0$, not directly on $\bx_t$, Diffusion Posterior Sampling (DPS)\cite{chung2023diffusion} estimates $\bx_0 \approx \E[\bx_0 \mid \bx_t]$ and adds a DC gradient with step-size $\zeta_t$:
\begin{subequations}\label{dps}
\begin{align}
\bx_{t-1} & \leftarrow \DiffusionStep(\bx_t, \bs_\theta^\star, t) \\
\bx_{t-1} & \leftarrow \bx_{t-1} - \zeta_t \nabla_{\bx_t} \| \by - \bcA(\E[\bx_0 \mid \bx_t]) \|_2^2 \label{eq:dps_dc}
\end{align}
\end{subequations}
The expectation is computed via Tweedie’s formula\cite{efron2011tweedies, robbins1992empirical}:
\begin{equation}\label{eq:tweedie}
\E[\bx_0 \mid \bx_t] = \frac{1}{\sqrt{\bar \alpha_t}} \left( \bx_t - \sqrt{1 - \bar \alpha_t} \bs_\theta^\star(\bx_t, t) \right)
\end{equation}
Here, $\bar \alpha_t~\in [0, 1]$ is the cumulative noise factor, which decreases monotonically with $t$\cite{ho2020denoising}.

In Latent-DPS (LDPS)~\cite{song2023solving, rout2023solving}, this update is applied in latent space, and the decoder $\cD$ is incorporated into the DC term in Eq.~\eqref{eq:dps_dc}:
\begin{equation}\label{eq:ldps}
\bz_{t-1} \leftarrow \bz_{t-1} - \zeta_t \nabla_{\bz_t} \| \by - \bcA(\cD(\E[\bz_0 \mid \bz_t])) \|_2^2
\end{equation}
Though introducing the overhead of backpropagation through $\cD$, LDPS enables efficient and flexible reconstruction from compressed representations using LDMs.

\vspace{-1em}\subsection{Longitudinally Accelerated Posterior Sampling}\label{sec:long_acc}
At a high level, we propose leveraging the prior scan to hot-start LDPS at an intermediate timestep $t_p < T$. In the following sections, we motivate this strategy, describe how to choose $t_p$ to best utilize the prior without over-biasing to it, and highlight the benefit of latent space for this prior injection.

We consider longitudinal reconstruction with a pre-trained complex LDM. This model defines a latent distribution at timestep $t=0$, denoted $p_0(\bz)$, assumed to approximate the true latent-image distribution. The posterior conditional distribution given measurements $\by$ is $p_0(\bz \mid \by)$. We denote the probability of sampling $\bz$, when starting the diffusion process from timestep $t$ with latent $\bz_t$, as $p_0(\bz \mid \bz_t)$, with unconditional sampling given by $p_0(\bz \mid \bz_T) = p_0(\bz)$.

In this setting, we assume access to a single prior scan, $\xp$, which shares contrast characteristics with a desired complex-valued follow-up scan, $\xgt$, with respective latent representations $\zp = \cE(\xp)$ and $\zgt = \cE(\xgt)$. For simplicity, in this section we assume that $\xp$ and $\xgt$ are aligned in phase, as $\cE$ has a complex input domain. In practice, $\xp$ will be magnitude only; this mismatch will be addressed in Section~\ref{sec:finding_good_init} by a phase initialization $\bphi\approx\angle\xgt$ such that $\zp = \cE(|\xp|e^{j \bphi})$.

\vspace{-0.75em}\subsubsection{Initializing Sampling in a Latent Manifold}\label{sec:latent_init}
Conventional diffusion sampling begins from pure Gaussian noise, $\bz_T \sim \cN(0, \mathbf{I})$, allowing unconditional sampling from the target distribution $p_0(\bz)$. Posterior sampling given $\by$ constrains the generative process towards the desired $\zgt$.
Ideally, $\zgt=\argmax{\bz}{p_0(\bz \mid \by)}$. But, as $\bcA$ becomes more ill-posed, the space of $p_0(\bz)$ consistent with $\by$ expands, hindering the exact recovery of $\zgt$.

To further constrain the probable output space of posterior sampling, we turn to the longitudinal setting. Assuming that $\xp$ and $\xgt$ share similar contrast and global structure, there is high likelihood that the scan pair share similar latent representations in $p_0(\bz)$, suggesting that $\zp$ could inform the sampling process of $\zgt$ via some soft constraint. Without explicitly learning this conditional relationship on a paired dataset, we propose to use $\zp$ only to initialize sampling. The key question then becomes how to do this effectively without over-biasing the reconstruction to $\xp$.

Consider first an ideal scenario: if $\zgt$ was accessible, we could directly project it to any intermediate diffusion timestep $t < T$ using\cite{ho2020denoising}:
\begin{equation}\label{eq:proj_zgt}
\zgt_t = \sqrt{\bar{\alpha}_t}\zgt + \sqrt{1-\bar{\alpha}_t}\epsilon \quad \epsilon \sim \mathcal{N}(0, \mathbf{I})
\end{equation}
Sampling from time $t<T$ initialized with $\zgt_t$ would inherently converge to $\zgt$ with higher probability than sampling from $T$ initiated by random noise $\bz_T$, as:
\begin{equation}\label{eq:zgt_proj_bound}
p_0(\bz_0 = \zgt \mid \bz_t = \zgt_t) > p_0(\bz_0 = \zgt \mid \bz_T) = p_0(\zgt).
\end{equation}

Building upon this intuition, we propose a similar initialization strategy using the prior. If we project $\zp$ to an intermediate timestep:
\begin{equation}\label{eq:proj_zp}
\zp_t = \sqrt{\bar{\alpha}_t},\zp + \sqrt{1-\bar{\alpha}_t}\epsilon, \quad \epsilon \sim \mathcal{N}(0, \mathbf{I}),
\end{equation}
then $t$ should be selected to maximize the likelihood of sampling $\zgt$, which we denote as $t_p$:
\begin{equation}\label{eq:opt_tp}
t_p = \argmax{t} {p_0(\zgt \mid \bz_t = \zp_t)}.
\end{equation}
In the worst case, setting $t_p = T$ reverts to traditional posterior sampling, maintaining the reconstruction quality of the unconditional case. However, if an intermediate $t_p < T$ improves the likelihood in Eq.~\eqref{eq:opt_tp}, $\zp_t$ becomes a better initialization for sampling towards $\zgt$, potentially enhancing reconstruction performance.

To select $t_p$ for a specific target-prior pair, we derive the following expression for Eq.~\eqref{eq:opt_tp}, assuming a Gaussian target latent space $\bz_0 \sim \mathcal{N}(0, \mathbf{I})$ (details in Supplementary~\ref{sec:tp_apdx}):
\begin{align}
\tp^\star & = \argmax{t}{\left\{\gamma_t + \frac{\bar{\alpha}_t}{2}\left[\frac{\|\zp\|_2^2}{2-\bar{\alpha}_t} - \frac{\|\zgt - \zp\|_2^2}{2(1-\bar{\alpha}_t)}\right]\right\}} \label{eq:opt_t_ex} \\
 & := \myoperatorname{TimeProject}(\zp, \zgt) \label{eq:timeproject}
\end{align}
Eq.~\eqref{eq:opt_t_ex} analytically relates the optimal initialization to a similarity measure between the scan pair, $\|\zgt - \zp\|_2^2$, additionally depending on noise parameters $\gamma_t$ and $\bar{\alpha}_t$. 

Intuitively, $\myoperatorname{TimeProject}$ selects a timestep where the diffusion noise level matches the latent discrepancy between scans. Under-selecting $\tp$ may overfit to the prior, while selection close to $T$ may discard useful prior information. Supplementary Fig.~\ref{fig:tp_sim} illustrates this trade-off: more similar scans favor smaller $\tp^\star$ to better leverage the prior, while dissimilar scans require larger $\tp^\star$ for flexibility. The figure also shows that Eq.~\eqref{eq:opt_t_ex} is concave in $t$ for fixed $\zp$ and $\zgt$, implying that an intermediate value is optimal.

In practice, we do not have access to $\zgt$ at inference, and the Gaussianity assumption on $\bz_0$ does not necessarily hold, thus we cannot directly apply Eq.~\eqref{eq:opt_t_ex}. In the next section, we thus propose an approximation for $\tp$, which is integrated into our full reconstruction method.

\vspace{-0.75em}\subsubsection{Finding a Good Initialization for LDPS}\label{sec:finding_good_init}
Hot-starting latent posterior sampling with the prior scan has two major hurdles:
\begin{enumerate}[leftmargin=*, label={(\arabic*)}]
    \item Prior scans are not phase-aligned (in our case, we assume access to magnitude-only DICOMs, i.e. $\xp$ is phase-less).
    \item The projection timepoint $\tp$ must consider the level of change between scans, including mis-registration and general scan variability.
\end{enumerate}

To address these, we introduce $\myoperatorname{AutoInit}$ (summarized in Algorithm~\ref{algo:autoinit}), which uses a preliminary reconstruction $\xinit$ generated through some fast reconstruction algorithm $\myoperatorname{InitReconAlg}(\cdot)$ to estimate both a projection timepoint $\tp$ and a phase map $\bphi\in\mathbb{R}^{N\times M}$. While any method can be used for $\myoperatorname{InitReconAlg}$, our chosen approach leverages a fast query of the trained LDM agnostic of $\xp$, further described in Section~\ref{sec:laps}.

To address (1), the initializing phase is extracted as $\bphi = \angle \xinit$. After registration to $\xinit$, the prior can then be updated as $|\xp|e^{j\bphi}$. For (2), we use $\zinit = \cE(\xinit)$ as a proxy for $\zgt$ when estimating $\tp$ via Eq.~\eqref{eq:timeproject}, using
\begin{equation}\label{eq:tp_approx}
\tpapprox = \myoperatorname{TimeProject}(\zp, \zinit).
\end{equation}
To account for the simplifying assumptions of Eq.~\eqref{eq:tp_approx}, we calibrate this estimate on a small validation set with empirically optimal $t_p$'s determined via grid search, fitting affine coefficients $v_p$ and $w_p$ such that the final projection timepoint is $t_p = v_p \cdot \tpapprox + w_p$ (see Supplementary~\ref{sec:auto_init_calib} and Fig.~\ref{fig:auto_tp} for details).

\vspace{-1em}\section{Methods}\label{sec:latent_recon}

\begin{figure*}[t]
  \begin{minipage}[t]{0.48\textwidth}
    \raggedright
    \begin{myalgorithm}{LAPS}\label{algo:laps}
      \begin{algorithmic}[1]
        \Require $T, \nstep, \, \nopt, \, \by, \, \bcA, \, \textcolor{ForestGreen}{\xp}, \, \, \bs_\theta, \cE, \cD$
        \State $\textcolor{ForestGreen}{\tp, \bphi = \myoperatorname{AutoInit}(\xp, \cE, \by, \bcA)}$
        \State $\tstep = \lfloor \frac{T}{\nstep} \rfloor$ \Comment{DDIM step size}
        \State $\cT = \{\tp, \tp - \tstep, \tp - 2\tstep, ..., 0\}$ \Comment{DDIM timesteps}
        \State $\beps \sim \cN(0, I)$
        \State $\bz_t = \bz_{\tp} = \sqrt{\bar \alpha_{\tp}} \cE(\textcolor{ForestGreen}{| \xp | \cdot e^{j \bphi }}) + \sqrt{1 - \bar \alpha_{\tp}} \beps$
        \For{$i = 1$ \textbf{to} $|\cT|-1$}
          \State $t, \tprev = \cT[i], \cT[i+1]$
          \State $\bz_{\tprev} = \DDIMStep(\bz_t, t, \tprev, \bs_\theta)$
          \For{$n = 1$ \textbf{to} $\nopt$} \Comment{Adam Optimization}
            \State $\hat \bz_0 = \E[\bz_0 \mid \bz_{\tprev}]$
            \State $\bz_{\tprev} = \bz_{\tprev} - \zeta_i \,\nabla_{\bz_{\tprev}}\,\| \by - A (\cD(\hat \bz_0) \|_2^2$
          \EndFor
        \EndFor
        \State $\hat{\bx} \leftarrow \text{CG}(\|\bcA(\bx) - \by\|_2^2,\;\; \bx_{\text{init}} = \cD(\bz_0))$ \Comment{Output DC}
        \State \Return $\hat{\bx}$
      \end{algorithmic}
    \end{myalgorithm}
  \end{minipage}
  \hfill
  \begin{minipage}[t]{0.48\textwidth}
    \raggedright
    \begin{myalgorithm}{CAPS}\label{algo:caps}
      \begin{algorithmic}[1]
        \Require $T, \nstep, \, \textcolor{BrickRed}{\tp}, \, \nopt, \, \by, \, \bcA, \, \bs_\theta, \cE, \cD$
        \State $\textcolor{BrickRed}{\xinit = \CG(\by, \bcA)}$
        \State $\tstep = \lfloor \frac{T}{\nstep} \rfloor$ \Comment{DDIM step size}
        \State $\cT = \{\tp, \tp - \tstep, \tp - 2\tstep, ..., 0\}$ \Comment{DDIM timesteps}
        \State $\beps \sim \cN(0, I)$
        \State $\bz_t = \bz_{\tp} = \sqrt{\bar \alpha_{\tp}} \cE(\textcolor{BrickRed}{\xinit}) + \sqrt{1 - \bar \alpha_{\tp}} \beps$
        \For{$i = 1$ \textbf{to} $|\cT|-1$}
          \State $t, \tprev = \cT[i], \cT[i+1]$
          \State $\bz_{\tprev} = \DDIMStep(\bz_t, t, \tprev, \bs_\theta)$
          \For{$n = 1$ \textbf{to} $\nopt$} \Comment{Adam Optimization}
            \State $\hat \bz_0 = \E[\bz_0 \mid \bz_{\tprev}]$
            \State $\bz_{\tprev} = \bz_{\tprev} - \zeta_i \,\nabla_{\bz_{\tprev}}\,\| \by - A (\cD(\hat \bz_0) \|_2^2$
          \EndFor
        \EndFor
        \State $\hat{\bx} \leftarrow \text{CG}(\|\bcA(\bx) - \by\|_2^2,\;\; \bx_{\text{init}} = \cD(\bz_0))$ \Comment{Output DC}
        \State \Return $\hat{\bx}$
      \end{algorithmic}
    \end{myalgorithm}
  \end{minipage}
  \vspace{-1.5em}
\end{figure*}
\begin{figure}[!h]
    \raggedright
    \begin{myalgorithm}{AutoInit}\label{algo:autoinit}
      \begin{algorithmic}[1]
        \Require $\xp, \, \cE, \, \by,\, \bcA$
        \Constant $\tpscale, \, \tpshift, \, \myoperatorname{InitReconAlg}$
        \State $\xinit = \myoperatorname{InitReconAlg}(\by, \bcA)$
        \State $\bphi = \angle \xinit$
        \State $\tilde \tp = \myoperatorname{TimeProject}(\cE(|\xp| \cdot e^{j\bphi}), \cE(\xinit))$
        \State $\tp = \tpscale \cdot \tilde \tp + \tpshift$
        \State \Return $\tp$, $\bphi$
      \end{algorithmic}
    \end{myalgorithm}
    \vspace{-1.5em}
\end{figure}

\begin{figure*}
    \centering
    \includegraphics[width=\linewidth]{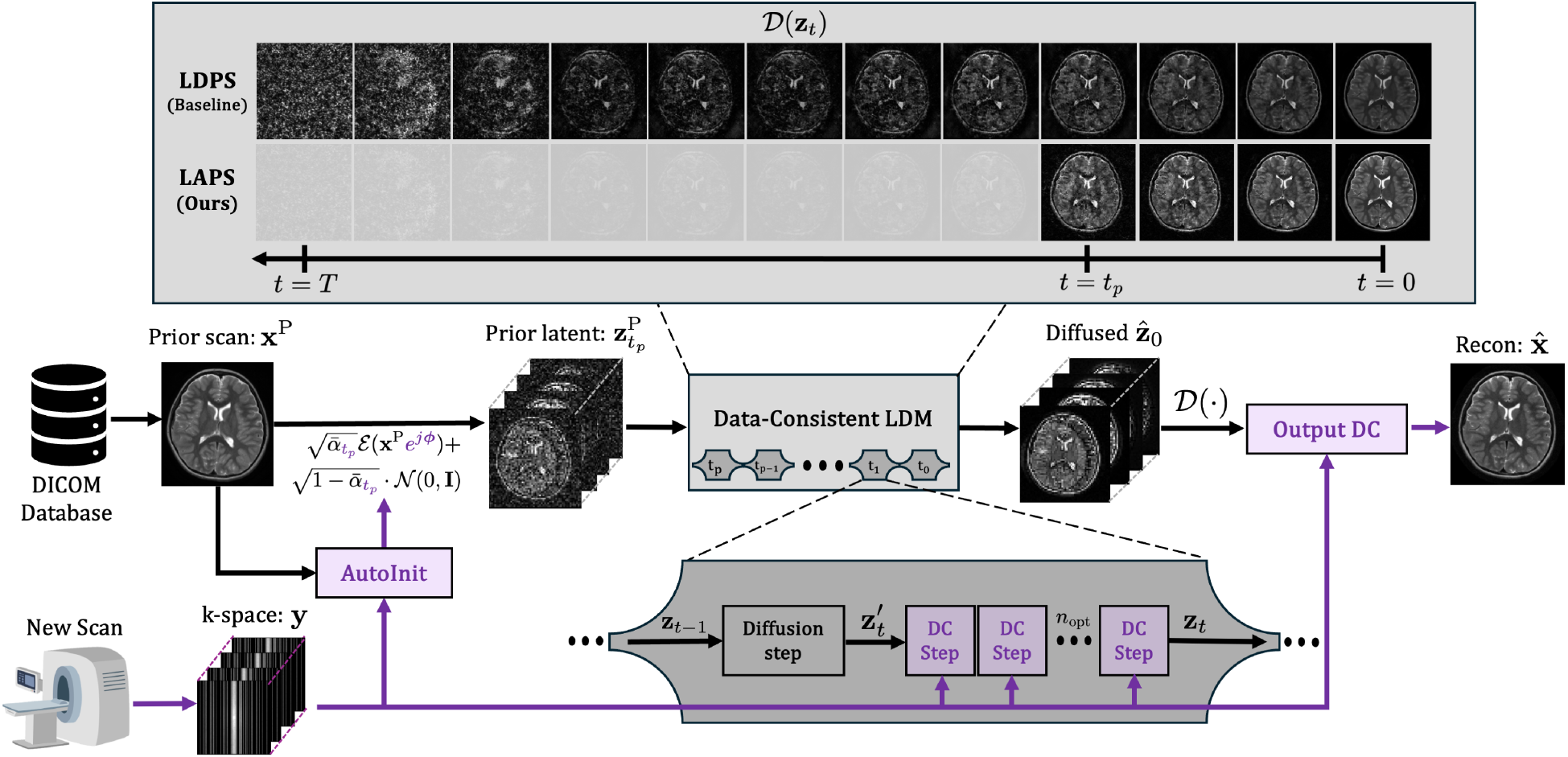}
    \caption{~Illustration of our proposed method, LAPS. From left to right: We begin with undersampled k-space from a new scan and retrieve the corresponding prior DICOM scan of the subject. Next, we estimate an initial phase $\bphi$ and find the optimal projection point $t_p$ using $\myoperatorname{AutoInit}$. We encode the phase-modulated prior and project it to $t_p$, which provides the initialization for our proposed LDM reconstruction that also incorporates the newly acquired k-space data. Finally, the decoded output of the diffusion process is passed through a few additional DC steps, refining the output image. Compared to LDPS (above), LAPS starts the diffusion process much earlier than $T$ via $\myoperatorname{AutoInit}$, runs $\nopt$ DC steps per iteration rather than 1, and enforces additional DC at the output.}
    \label{fig:method}
\end{figure*}

Building on the framework introduced in the previous sections, we now present the complete reconstruction method. We first pretrain an LDM on unpaired complex images. Then, we use the prior-based initialization scheme introduced in Section~\ref{sec:long_acc} in conjunction with a modified LDPS\cite{rout2023solving} approach. Specifically, instead of using a single latent DC gradient step as in Eq.~\eqref{eq:ldps}, we use $\nopt>1$ steps. This iterative correction improves performance by perturbing each $z_t$ closer to the manifold of data consistent latents. To control the update size adaptively, we use an Adam optimizer\cite{KingBa15}.

To enable faster reconstruction with fewer diffusion steps, we replace the standard DDPM sampling in LDPS with DDIM. Finally, since the learned decoder $\cD$ is lossy, we include additional image-domain DC steps at the output of latent sampling. 

We name this reconstruction method \textbf{L}ongitudinally \textbf{A}ccelerated \textbf{P}osterior \textbf{S}ampling (LAPS), illustrated in Fig.~\ref{fig:method} and summarized in Algorithm~\ref{algo:laps}. To improve on a single random initialization, our final reconstruction averages over $n_{\rm{avg}}$  samples generated in parallel, known in diffusion literature to further improve image quality \cite{jalal2021robust}.

As a variant of our reconstruction framework that does not use the prior scan, we introduce \textbf{C}onjugate Gradient-Initialized \textbf{A}ccelerated \textbf{P}osterior \textbf{S}ampling (CAPS). Instead of the prior, it initializes latent sampling from a CG reconstruction projected to a fixed $\tp$, then follows the same steps as LAPS (see Algorithm~\ref{algo:caps} for direct comparisons). In this work, CAPS serves as a strong prior-free baseline, matching or outperforming other latent posterior sampling methods such as PLDS\cite{rout2023solving} in our experiments (see Supplementary Fig.~\ref{fig:latent_sampling_methods}), while enabling faster reconstruction. The comparison between CAPS and LAPS isolates the added benefit of incorporating longitudinal information into reconstruction.

\begin{table}[ht]
\centering

\begin{subtable}[t]{0.5\textwidth}
\centering
\bitsmall
\begin{tabular*}{0.98\linewidth}{@{\extracolsep{\fill}} l  r  rr  rr}
\toprule
\textbf{Category} & \textbf{Total} & \multicolumn{2}{c}
{\textbf{Train}} & \multicolumn{2}{c}{\textbf{Test}} \\
\toprule
& \textbf{All}   & \textbf{All} & {Pair} & \ \ \textbf{Pair} \\
\midrule
\textbf{Subjects} & 53   & 47   & 25   & 6   \\
\textbf{Scanners} & 13   & 13   & 7    & 3   \\
\textbf{Scans}    & 232  & 209  & 81   & 21  \\
\textbf{2D slices}& 34970 & 32600 & 9968 & 1970\\
\bottomrule
\end{tabular*}
\caption{General statistics of SLAM dataset.}
\label{tab:slam_overview}
\end{subtable}

\centering
\begin{subtable}[t]{0.5\textwidth}
\centering
\bitsmall
\vspace{1ex}
\begin{tabular*}{0.98\linewidth}{@{\extracolsep{\fill}} l  r  rr  rr}
\toprule
\textbf{Protocol} & \textbf{Total} & \multicolumn{2}{c}
{\textbf{Train}} & \multicolumn{2}{c}{\textbf{Test}} \\
\toprule
& \textbf{All}   & \textbf{All} & {Pair} & \ \textbf{Pair} \\
\midrule
\textbf{Brain Tumor} & 106 & 90 & 36 & 14 \\ 
\textbf{Brain \& Pituitary} & 29 & 29 & 14 &- \\ 
\textbf{Glioma/Radiation} & 23 & 23 & - & - \\ 
\textbf{Brain \& Orbits} & 19 & 19 & 12 & - \\ 
\textbf{Stroke} & 9 & 9 & - & - \\ 
\textbf{Ped Brain} & 13 & 8 & 8 & 5 \\ 
\textbf{ROSA} & 7 & 7 & 6 & - \\ 
\textbf{Brain Vascular} & 8 & 6 & 3 & 2 \\ 
\textbf{Unk. Brain Mass} & 4 & 4 & - & - \\ 
\textbf{Seizures} & 4 & 4 & 2 &  - \\ 
\textbf{Epilepsy} & 3 & 3 & - & - \\ 
\textbf{Other} & 7 & 7 & - & - \\ 
\bottomrule
\end{tabular*}
\caption{Distribution of scans by protocol.}
\label{tab:slam_protocols}
\end{subtable}

\begin{subtable}[t]{0.5\textwidth}
\centering
\bitsmall
\vspace{1ex}
\begin{tabular*}{0.98\linewidth}{@{\extracolsep{\fill}} l  r  rr  r}
\toprule
\textbf{Scan Type} & \textbf{Total} & \multicolumn{2}{c}
{\textbf{Train}} & \textbf{Test} \\
\toprule
& \textbf{All}   & \textbf{All} & {Pair} & \ \textbf{Pair} \\
\midrule
\textbf{$T_2$ (2D)} & 83 & 72 & 40 & 11 \\ 
\textbf{$T_2$ FLAIR (3D)} & 40 & 37 & 17 & 3 \\ 
\textbf{$T_1$ BRAVO (3D)} & 24 & 23 & 7 & 1 \\ 
\textbf{$T_1$ BRAVO+GAD (3D)} & 23 & 19 & 9 & 4 \\ 
\textbf{$T_2$ (3D)} & 21 & 18 & 4 & 3 \\ 
\textbf{$T_1$ FSE (3D)} & 15 & 15 & - & - \\ 
\textbf{$T_1$ CUBE (3D)} & 11 & 10 & 2 & 1 \\ 
\textbf{$T_2$ FLAIR (2D)} & 7 & 7 & 2 & - \\ 
\textbf{$T_1$ SPGR (3D)} & 5 & 5 & - & - \\ 
\textbf{$T_1$ FLAIR (2D)} & 3 & 3 & - & - \\ 
\bottomrule
\end{tabular*}
\caption{SLAM scans by scan type.}
\label{tab:slam_scan_types}
\end{subtable}

\vspace{1ex}
\begin{subtable}[t]{0.225\textwidth}
\centering
\begin{tabular}{l r r}
\toprule
\textbf{Interval} & \textbf{Train} & \textbf{Test} \\
\midrule
\textbf{$<$ 1 mo.} & 2 & 1 \\ 
\textbf{1--4 mo.} & 11 & 3 \\ 
\textbf{4--8 mo.} & 9 & 2 \\ 
\textbf{$>$ 8 mo.} & 5 & -- \\ 
\textbf{N/A} & 20 & -- \\ 
\bottomrule
\end{tabular}
\caption{Time between scan sessions.}
\label{tab:slam_times}
\end{subtable}
\hfill
\begin{subtable}[t]{0.23\textwidth}
\centering
\begin{tabular}{l r r}
\toprule
\textbf{Change} & \textbf{Train} & \textbf{Test} \\
\midrule
\textbf{Low}  & 12 & 2 \\ 
\textbf{Med.} & 8 & 3 \\ 
\textbf{High} & 7 & 1 \\ 
\textbf{N/A}  & 20 & -- \\ 
& & \\
\bottomrule
\end{tabular}
\caption{Radiologist change ratings.}
\label{tab:slam_change}
\end{subtable}
\caption{~Summary statistics of scans in SLAM dataset. Tables~(\subref{tab:slam_overview}), (\subref{tab:slam_protocols}) and (\subref{tab:slam_scan_types}), list the total number of scans with raw k-space data (All), which can be used for training the LDM, and the subset that are paired, i.e., include a matching prior scan (Pair), which can be used for validating our method. SLAM was split by subject into train and test datasets, with test set scans filtered only to paired examples required for evaluation. Tables~(\subref{tab:slam_protocols}) and (\subref{tab:slam_scan_types}) list quantities by scan counts, while (\subref{tab:slam_times}) and (\subref{tab:slam_change}) list by subject.}
\label{tab:slam}
\end{table}

\vspace{-1em}\subsection{Stanford Longitudinally Accelerated MRI (SLAM) Dataset}
A major barrier to developing and evaluating longitudinal MRI reconstruction methods is the lack of available datasets suited for the task. To fill this gap, we collaborated with Stanford Hospital to curate a dataset containing prior DICOM scans and follow-up scans with k-space across multiple neuroimaging protocols. We call this the \textbf{S}tanford \textbf{L}ongitudinally \textbf{A}ccelerated \textbf{M}RI (SLAM) Dataset, which spans a variety of scan types, contrasts, intervals, and pathologies.

The LAPS framework has different data requirements for training and testing: training the LDM requires a dataset of complex-valued reconstructions (not longitudinally paired), whereas evaluation requires paired data consisting of a prior DICOM scan and the raw k-space of a follow-up scan. SLAM supports both: it provides raw k-space with corresponding complex reconstructions for LDM training, and many scans also include prior DICOMs for longitudinal evaluation.

In this paper, we present results based on SLAM as of writing. Tables~\ref{tab:slam_overview}, \ref{tab:slam_protocols}, and \ref{tab:slam_scan_types} summarize its size, originating protocols, and scan types.  Tables~\ref{tab:slam_times} and \ref{tab:slam_change} list the elapsed intervals between scans and radiologist-graded change levels, demonstrating range in the variability between scan sessions in the dataset. All data were acquired on 3.0 T GE systems and split by subject into training and testing sets.

Acquired in typical clinical settings, most scans were collected with 24–48-channel head coils and mild undersampling (R=2 for 2D acquisitions; R=2–5 for 3D cases, of which a subset had variable density sampling for CS reconstruction). To get complex-valued images of the follow-up k-space for network training, we implemented our own reconstruction pipeline on the raw multi-coil k-space. First, we trimmed the raw k-space to a maximum matrix size of $256\times256$ for consistency and network compatibility, followed by SVD-based coil compression to retain $95\%$ of the SVD energy along the coil dimension. We then performed an L1-wavelet CS~\cite{lustig2007sparse} reconstruction to get complex-valued images, leveraging sensitivity maps estimated from fully sampled center regions via ESPIRiT~\cite{uecker2014espirit}. 
 We excluded scans with visible artifacts (e.g., motion, low SNR) or poor reconstruction quality, yielding a final dataset of 232 unique scans with complex-valued follow-up reconstructions. With our train/test partition, this generated 209 complex scans for network training, of which 81 scans had paired magnitude DICOM scans from a previous session.

For this preliminary study we focus on 2D reconstructions: as such, 3D acquisitions were converted to 2D slices via an inverse FFT along the readout dimension. For each slice, we saved the full k-space, acquisition mask, and our reconstruction, enabling further undersampling retrospectively (see examples in Supplementary Fig.~\ref{fig:retrospective_undersampling}).

For scans with prior DICOM images, we registered across sessions using an affine rigid transformation provided by the Advanced Normalization Tools (ANTs) library\cite{avants2009advanced, avants2008symmetric}. Though longitudinal reconstruction methods depend on previous scans which may not be registered at inference, we assume a rough affine registration can first be completed with an initial reconstruction, which we already require with LAPS for phase initialization. Unless otherwise noted, we assume scans are registered in our analysis, but we additionally compare performance under controlled mis-registration settings.

We are actively collecting new data for SLAM, as we plan to expand our current public release to enable development on more complex longitudinal reconstruction frameworks in the future.

\vspace{-1em}\subsection{Model Evaluation}

To evaluate our proposed method, we designed several experiments involving various longitudinal and undersampling conditions. The following section describes implementation details of our method and selected baselines, which includes both classical and learning-based approaches, with and without longitudinal priors. 

\vspace{-0.75em}\subsubsection{LAPS: Our Proposed Method}\label{sec:laps}
We first trained a 2D LDM on all of the follow-up reconstructions in the SLAM training set (32600 complex-valued slices), as well as FastMRI brain~\cite{fastmri} (13000 $T_2$ and 5300 $T_1$ complex-valued slices). We additionally included the approximately 10000 slices of magnitude-valued DICOMs of prior scans in SLAM for model training. To use DICOMs for complex-valued training, we added a random quadratic spatially-varying phase as an approximate model for the RF phase profiles resulting on coil-combined images, assuming phase examples from susceptibility or static field inhomogeneity are otherwise well represented by the true complex reconstructions within the dataset. For the first stage of LDM training, we fine-tuned Med-VAE\cite{varma2025medvae}, a generalized autoencoder for medical images, on our training data, selecting $K=4$ and $C=4$ for Eq.~\eqref{eq:vae}. Since Med-VAE originally operates on magnitude images, to handle our complex data, we modified the architecture to take a 2-channel input by replicating the weights of the first layer per channel, and then stacked the real and imaginary components of the complex input. Fine-tuning followed the original implementation\cite{varma2025medvae}, trained across 5 NVIDIA A6000 GPUs for 4 days.

In our second stage, we trained the LDM ($\bs_\theta$). Specifically, we developed a modified denoising architecture from the UNet of Stable Diffusion (SD) 1.5\cite{rombach2022highresolution}, a high resolution text-to-image LDM. Since our approach does not use text conditioning, we replaced the cross-attention layers of the SD UNet architecture with self-attention layers, effectively distilling out the learned conditioning of the model on text. We then fine-tuned the UNet weights on our training datasets encoded with the fine-tuned Med-VAE, with the same compute and duration as VAE training. 

For $\myoperatorname{InitReconAlg}$ to get $\xinit$, we used our same reconstruction network with a highly accelerated CAPS reconstruction, fixing $\tp=200$, $\nopt=5$, $\nstep=80$, and $\navg=1$, which takes approximately 1/10 as long as the full LAPS reconstruction. Using this $\xinit$, we calibrated the $\tp$ affine transform used in $\myoperatorname{AutoInit}$, yielding $\tpscale=1.55, \tpshift=-350$ (additional detail in Supplementary~\ref{sec:auto_init_calib_implementation}).

For the remaining hyperparameters, we found $\nopt = 10$ to perform best with $n_{\rm{step}}=100$ total DDIM steps (see Supplementary Fig.~\ref{fig:nopt}), followed by 6 CG steps for the output DC. We report final reconstructions as the average over $n_{\rm{avg}}=4$ samples.

\vspace{-0.75em}\subsubsection{Baseline Reconstruction Methods}\label{sec:baselines}
First, we compare our approach with other longitudinal reconstruction methods. For fair comparison to LAPS, we only consider unsupervised methods which do not require longitudinally-paired training datasets, relying on the prior DICOM $\xp$ only at inference:

\begin{enumerate}[leftmargin=*]
    \item \textbf{LACS} \cite{weizman2015compressed}: A compressed sensing method combining wavelet sparsity with similarity to $\xp$.
    \item \textbf{NERP} \cite{shen2022nerp}: Uses an INR trained to represent $\xp$, fine-tuned via DC with the follow-up k-space. 
\end{enumerate}

\noindent To ablate the utility of using longitudinal information in reconstruction, we compare with other DL methods which are agnostic of prior scans: 

\begin{enumerate}[leftmargin=*]
    \setcounter{enumi}{2}
    \item \textbf{MODL} \cite{aggarwal2018modl}: A DL framework unrolling CG DC with CNN-based denoising, trained end-to-end per acceleration rate with supervised references. 
    \item \textbf{AdaDiff} \cite{dar2022adaptive}: An image-space diffusion-based reconstruction method, implemented using the official open-source code and trained on the same dataset as LAPS. AdaDiff benchmarks our LDM approach against a state-of-the-art image-domain approach.
    \item \textbf{Conjugate Gradient-Initialized LDPS (CAPS)}: As described in Section~\ref{sec:latent_recon}. Except for fixing $t_p=200$, all hyperparameters match those used in LAPS. 
\end{enumerate}

\noindent Additional details and hyperparameters for baseline implementations are provided in Supplementary~\ref{sec:baseline_apdx}.

\vspace{-0.75em}\subsubsection{Metrics}\label{sec:eval_metrics}
We assess performance of our method and baselines on the SLAM test partition (21 scans from 6 subjects, per Table~\ref{tab:slam_overview}). We evaluate reconstructions on retrospectively undersampled k-space at various rates with variable density sampling (see Supplementary~\ref{sec:undersampling} for our retrospective undersampling strategy). 2D or 3D acquisitions with a single undersampling dimension are evaluated with 1D retrospective undersampling, whereas 3D acquisitions with two undersampling dimensions are evaluated only with 2D retrospective undersampling. For scans that were fully sampled in the original acquisition, we generate both 1D and 2D retrospective undersampling patterns, allowing evaluation under both cases. All volumetric reconstructions are completed slice by slice. For evaluation, we include only 15 evenly spaced slices per volume to avoid biasing to scans with more slices, resulting in evaluation over 225 slices with 1D undersampling and 150 slices for 2D undersampling. All retrospective coil-compressed k-space retains a fully sampled center, included in the reported undersampling rate $R$, used for ESPIRiT calibration prior to reconstruction. 

To assess global reconstruction quality, we compute PSNR and SSIM over all test set slices. Taken standalone, these metrics can be misleading in longitudinal reconstruction, where much of the anatomy may remain unchanged, potentially masking errors in regions with differences. To address this, we also evaluate performance in local image regions classified by similarity between scan timepoints. Per prior-target scan pair, we divide pixel space into $32 \times 32$ patches with $50\%$ overlap and compute cosine similarity between corresponding patch pairs. We then aggregate all patches across the test set, bin by cosine similarity percentile, and report average reconstruction PSNR and SSIM over all patches per bin.

\vspace{-1em}\section{Results}\label{sec:results}
\vspace{-1em}\subsection{Performance vs. Acceleration}
\begin{figure*}[ht]    
    \centering
    \includegraphics[width=0.68\linewidth]{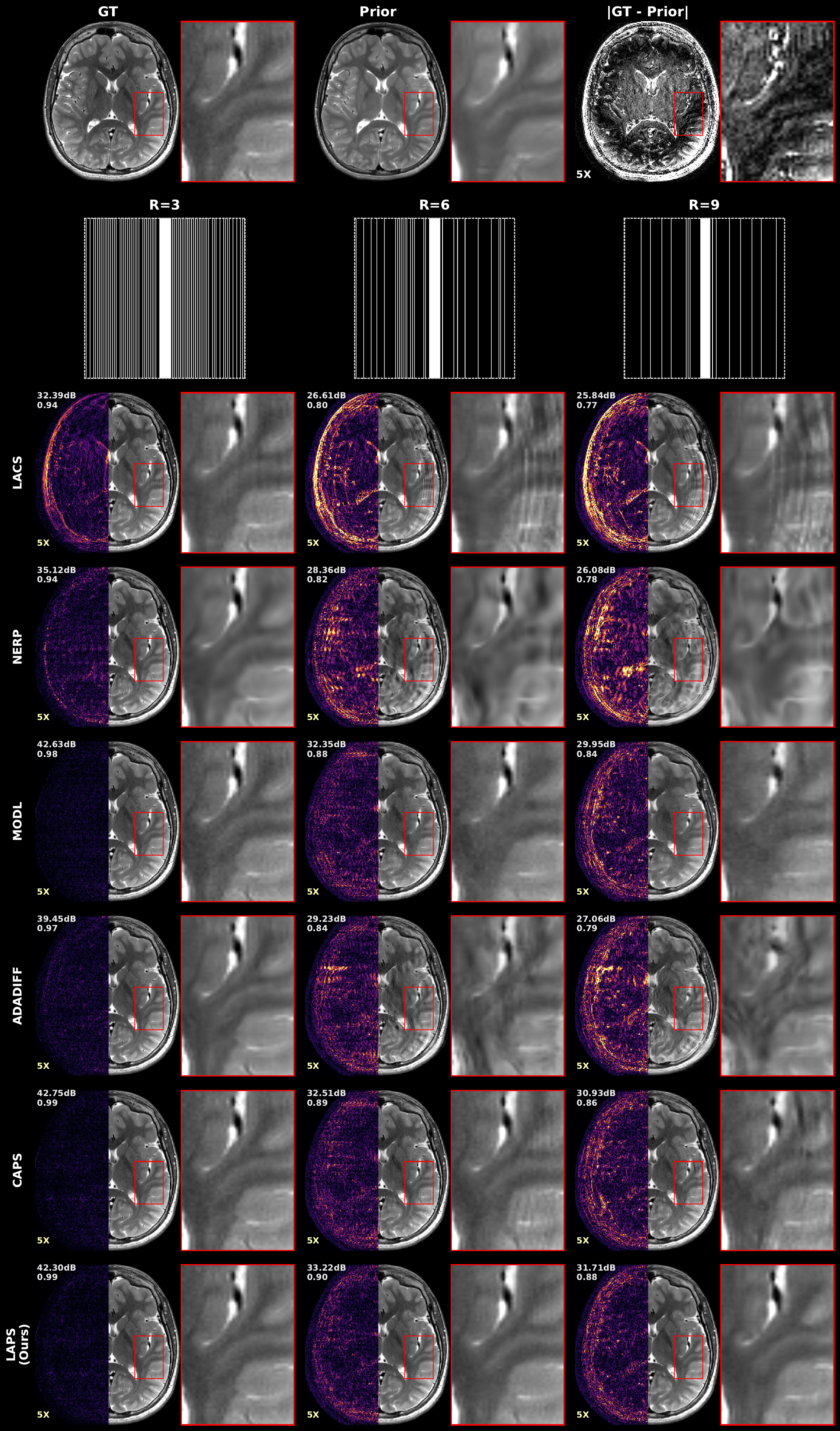}
    \vspace{-1em}
    \caption{~Comparison of reconstruction methods at varying 1D acceleration rates in a case with minimal longitudinal change.
    Axial $T_2$-weighted images at the level of the basal ganglia and insula in a patient with an incidentally discovered sub-centimeter cerebellar lesion (not shown), presumed to be a low-grade tumor. The lesion and other brain regions remained stable between the two exams performed six months apart.
    The first row shows, from left to right, the target (ground truth) image, the prior scan, and the amplified difference between them. The second row displays the undersampling masks used for reconstruction at acceleration rates of $R = 3$, $6$, and $9$.
    Subsequent rows present reconstructions from different methods, with our proposed method (LAPS) shown in the last row. Each image is split vertically: the left half shows the error map (amplified $\times5$) of the corresponding left half of the reconstruction, while the right half displays the actual reconstructed image. A zoomed-in region is shown to the right of each reconstruction to highlight fine details and local differences. Finally, the PSNR and SSIM of each reconstruction are indicated at the top left corner.}
    \label{fig:recon_accel_normal}
\end{figure*}

We first evaluate reconstruction performance on cases with minimal change between scan timepoints. Fig.~\ref{fig:recon_accel_normal} shows an axial $T_2$-weighted example across methods and 1D acceleration rates. At low acceleration ($R=3$), all methods perform similarly, although LACS shows slightly more aliasing and noise. As acceleration increases, errors increase across all methods but remain consistently lower for LAPS. At $R=9$, AdaDiff and NERP show loss of structural fidelity, and LACS errors resemble the prior–target difference map. While CAPS displays aliasing artifacts, LAPS retains a relatively clean reconstruction, showing noise-like errors.

\begin{figure*}[ht]
    \centering
    \includegraphics[width=0.98\linewidth]{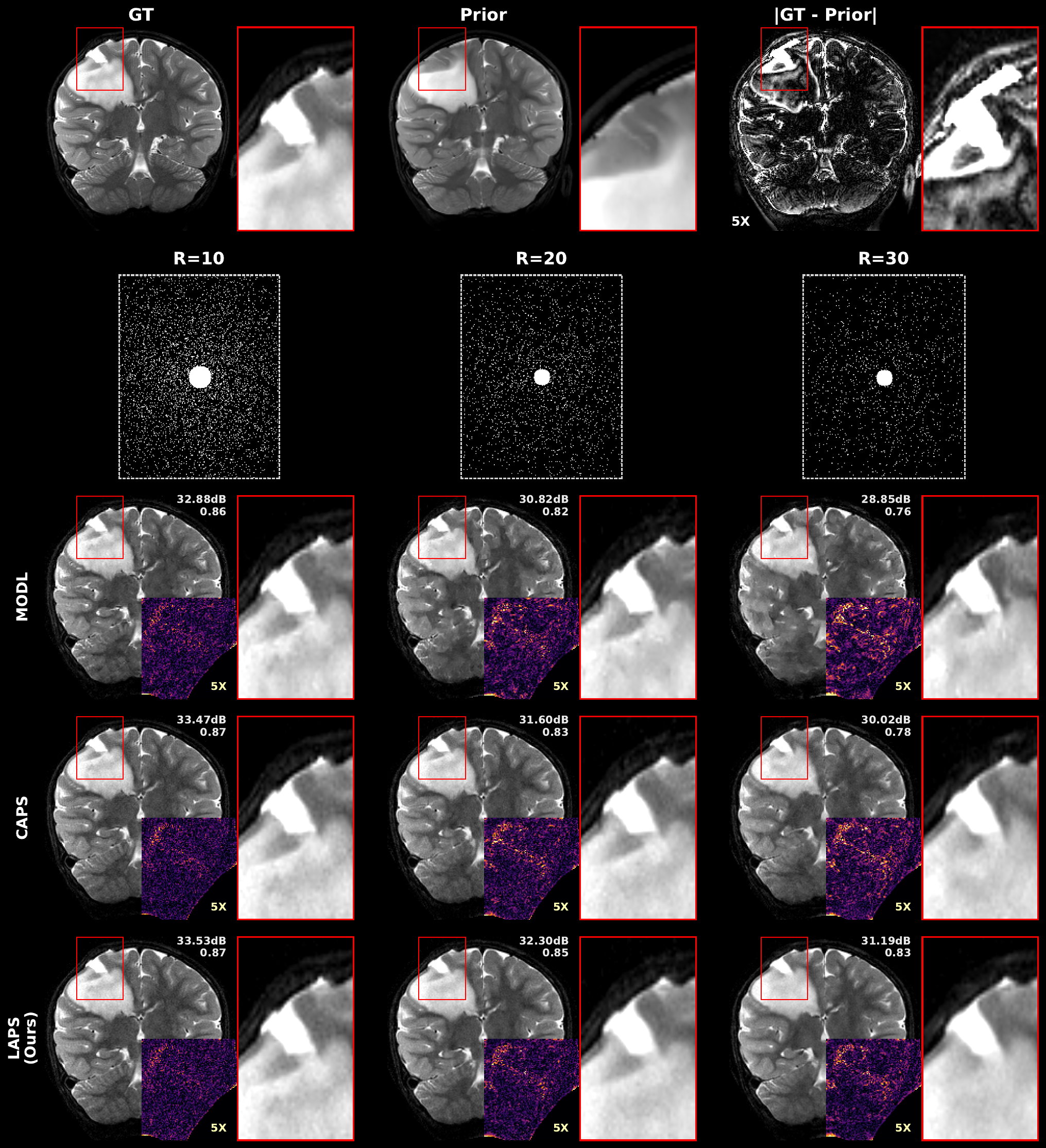}
    \vspace{-1em}
    \caption{~Comparison of reconstruction methods across varying 2D acceleration rates in a case with visible longitudinal changes.
    Coronal $T_2$-weighted images from a pediatric brain scan with a low-grade glioma. The initial scan (prior) shows a 6 cm $T_2$-hyperintense mass in the right frontal and parietal lobes. In the follow-up scan (target), the mass has been biopsied, with the biopsy cavity clearly visible.
    The first row displays (left to right): the follow-up scan (target), the prior scan, and an amplified difference image highlighting changes between them.
    The second row shows the undersampling masks used in the experiments, corresponding to acceleration rates of $R = 10$, $20$, and $30$.
    The subsequent rows present reconstructed images from various methods, with LAPS shown in the bottom. The PSNR and SSIM of each reconstruction are indicated at the top right corner.
    A zoomed-in region around the biopsy cavity is used to assess each method’s ability to accurately reconstruct pathological tissue and capture deviations over time.}
    \label{fig:recon_accel_abnormal}
\end{figure*}

Fig.~\ref{fig:recon_accel_abnormal} shows a longitudinal case of a patient with low-grade glioma, retrospectively evaluated at various 2D acceleration rates for the top-performing models: LAPS, CAPS, and MODL (a similar figure with all methods is provided in Supplementary Fig.~\ref{fig:recon_accel_abnormal_full}). The prior exam shows a $T_2$-hyperintense mass in the right frontal and parietal lobes. The mass was biopsied, and the resulting biopsy cavity is visible in the follow-up scan, creating a noticeable local change.

At $R=10$, all methods recover the overall structure and clearly depict the cavity. At $R=20$ MODL begins to exhibit undersampling artifacts and blurring, while CAPS show emerging structural distortions. As the acceleration increases, MODL loses white/gray matter detail, but the cavity remains visible.

At $R=30$, all methods show some degradation. LAPS demonstrates some blurring at the base of the biopsy cavity but retains lower levels of aliasing artifacts and noise than other methods.

\begin{figure*}
  \centering
  \begin{subfigure}[t]{0.4\textwidth}
    \centering
    \includegraphics[width=\linewidth]{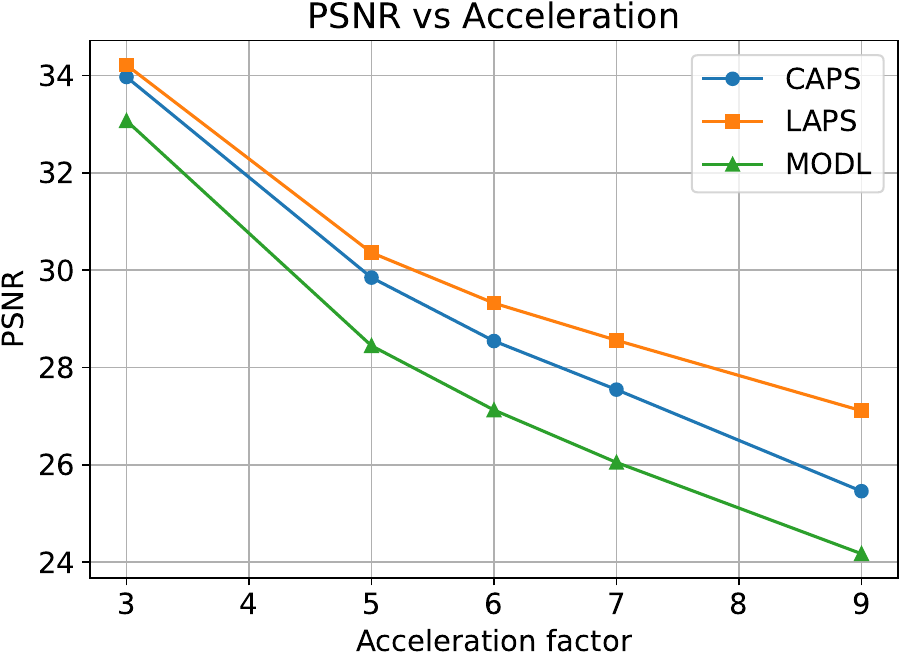}
    \caption{Global PSNR vs. 1D acceleration.}
    \label{fig:metrics_line_1d_psnr}
  \end{subfigure}%
  \hspace{1em}
  \begin{subfigure}[t]{0.4\textwidth}
    \centering
    \includegraphics[width=\linewidth]{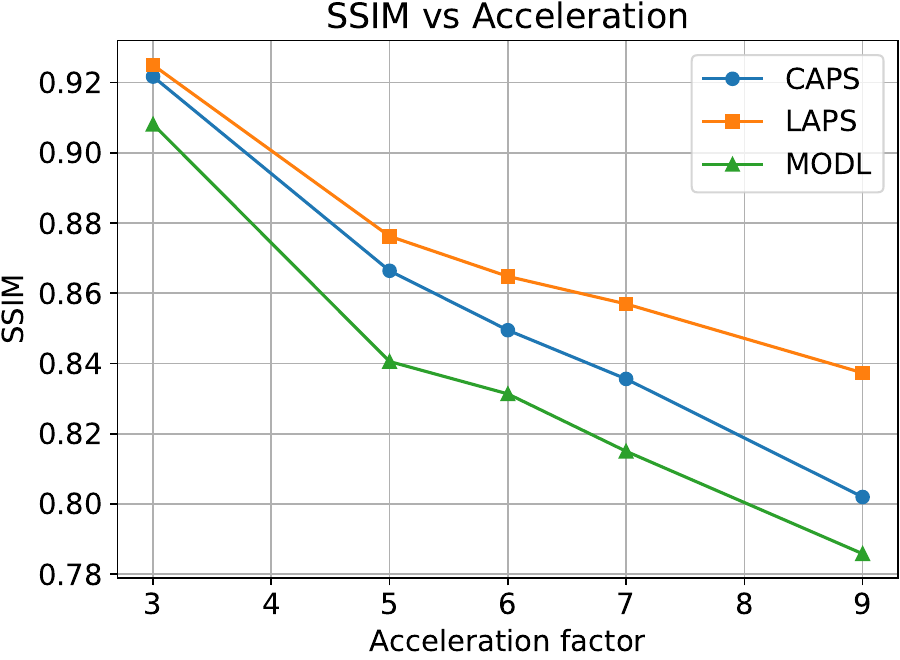}
    \caption{Global SSIM vs. 1D acceleration.}
    \label{fig:metrics_line_1d_ssim}
  \end{subfigure}%
  \vspace{1ex}
  
  \begin{subfigure}[t]{0.4\textwidth}
    \centering
    \includegraphics[width=\linewidth]{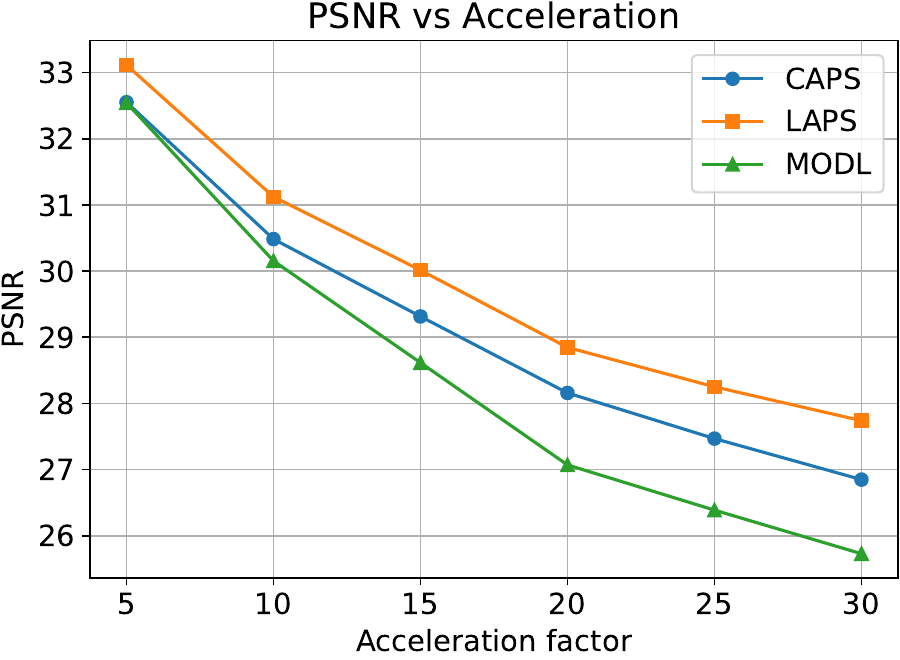}
    \caption{Global PSNR vs. 2D acceleration.}
    \label{fig:metrics_line_2d_psnr}
  \end{subfigure}%
  \hspace{1em}
  \begin{subfigure}[t]{0.4\textwidth}
    \centering
    \includegraphics[width=\linewidth]{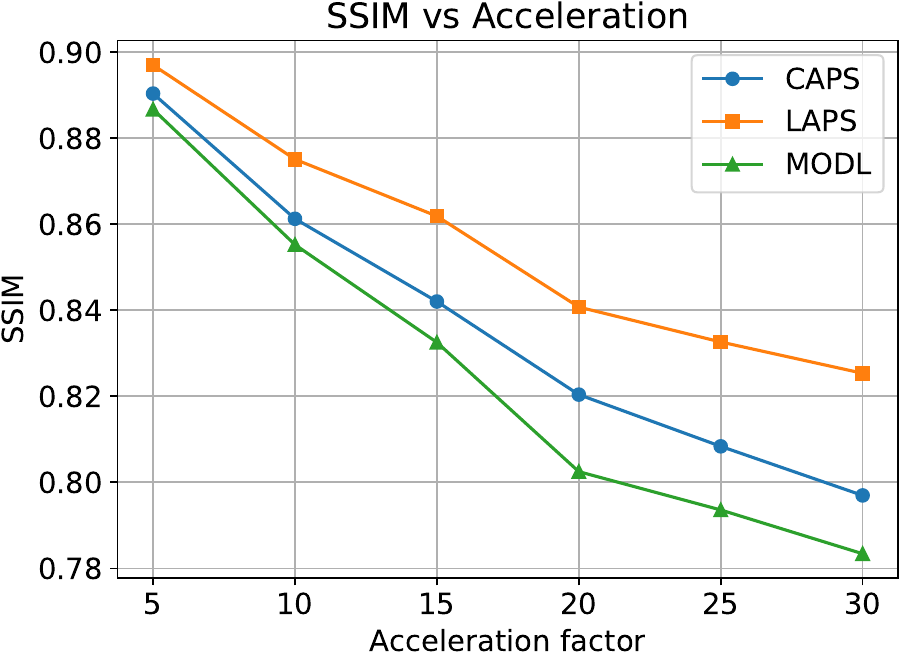}
    \caption{Global SSIM vs. 2D acceleration.}
    \label{fig:metrics_line_2d_ssim}
  \end{subfigure}%
  \vspace{1ex}

  \caption{~Mean global performance metrics across various 1D and 2D accelerations. For brevity, performance is only shown for LAPS with $\myoperatorname{AutoInit}$ and the top 2 performing baselines (CAPS and MODL).}
  \label{fig:metrics_line}
\end{figure*}

Tables~\ref{tab:metrics_1d} and \ref{tab:metrics_2d} report reconstruction metrics on the test set for 1D and 2D undersampling. LAPS consistently achieves the highest mean PSNR and SSIM across all acceleration rates, except for SSIM at $R=3$, where AdaDiff achieves a slightly higher value. At $R=6$, LAPS improves PSNR over CAPS by 0.7dB, increasing to 1.6dB at $R=9$. Similar trends are observed in 2D undersampling. Fig.~\ref{fig:metrics_line} shows plots of the PSNR and SSIM as a function of $R$ for the different methods, where the quantitative performance of LAPS at R=30 with 2D undersampling is visibly equal or better to performance of CAPS at about R=23 in PSNR and SSIM. In the 1D undersampling case, LAPS at R=9 is quantitatively comparable to CAPS at R=7.

\begin{figure*}[htbp]
  \centering
  \begin{subfigure}[t]{0.48\textwidth}
    \centering
    \includegraphics[width=\linewidth]{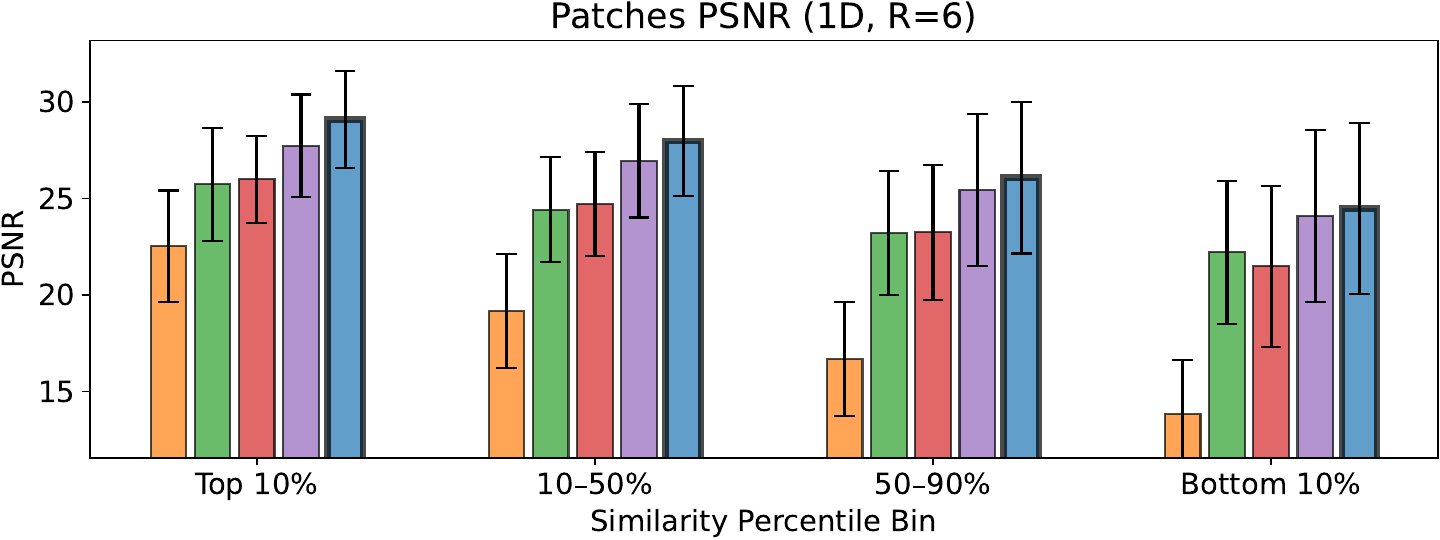}
    \caption{Patch-based PSNR for 1D acceleration at $R=6$.}
    \label{fig:patch_metrics_r6_psnr}
  \end{subfigure}%
  \hfill
  \begin{subfigure}[t]{0.48\textwidth}
    \centering
    \includegraphics[width=\linewidth]{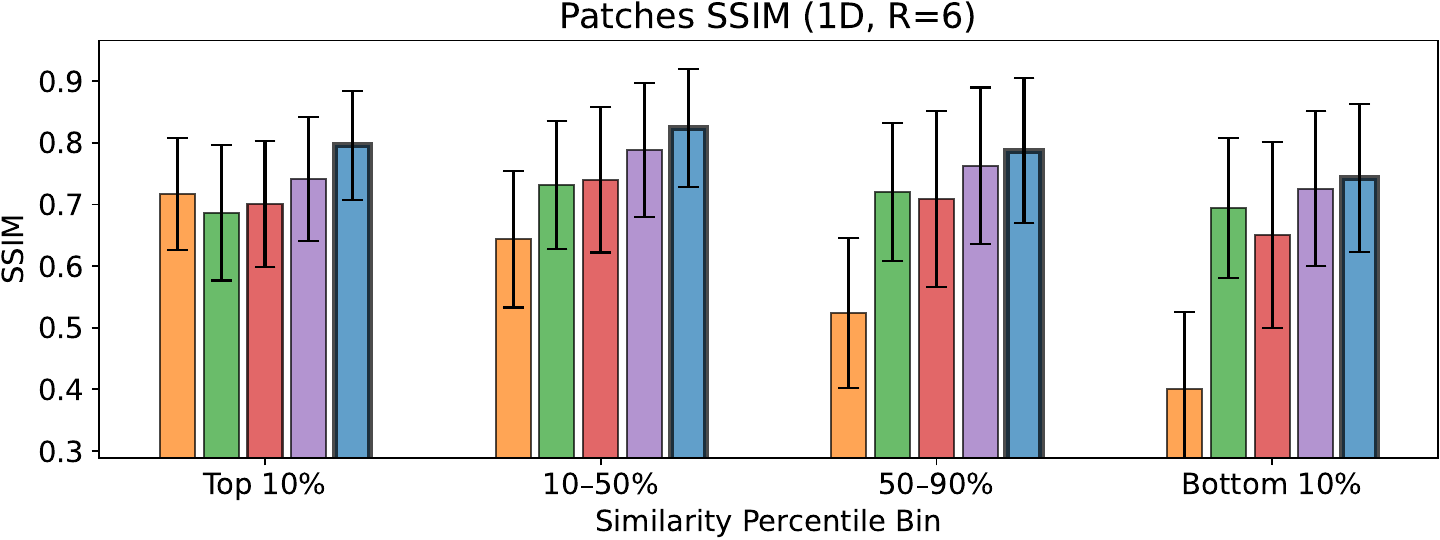}
    \caption{Patch-based SSIM for 1D acceleration at $R=6$.}
    \label{fig:patch_metrics_r6_ssim}
  \end{subfigure}
  
  \vspace{1ex}

  \begin{subfigure}[t]{0.48\textwidth}
    \centering
    \includegraphics[width=\linewidth]{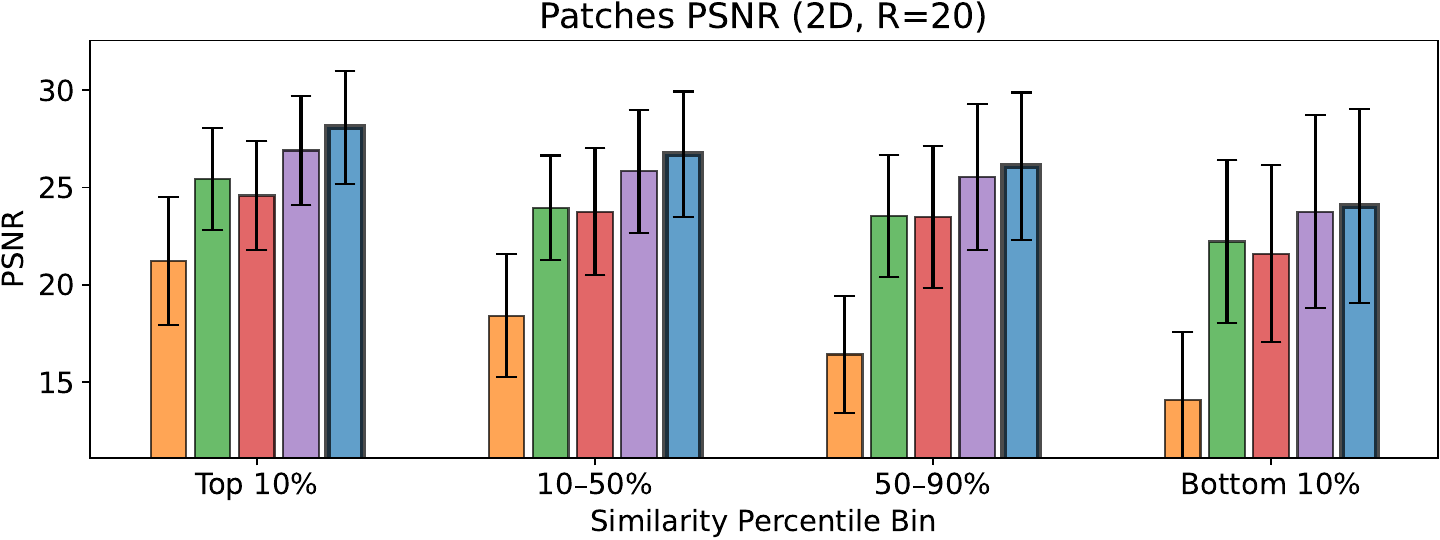}
    \caption{Patch-based PSNR for 2D acceleration at $R=20$.}
    \label{fig:patch_metrics_r20_psnr}
  \end{subfigure}%
  \hfill
  \begin{subfigure}[t]{0.48\textwidth}
    \centering
    \includegraphics[width=\linewidth]{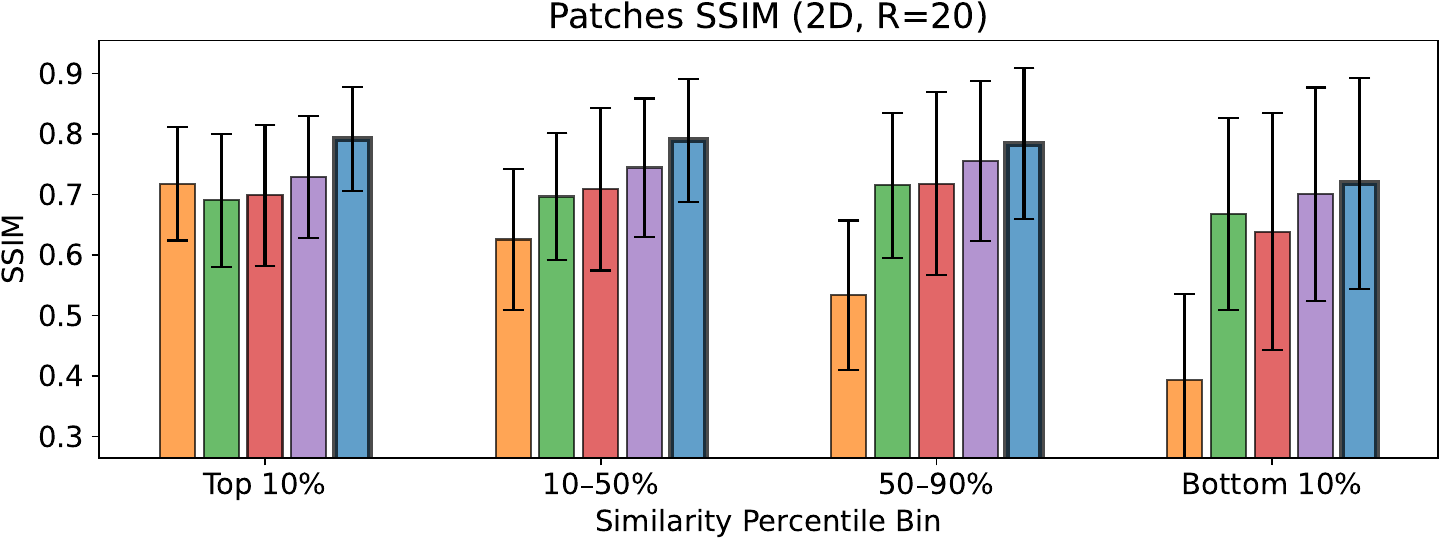}
    \caption{Patch-based SSIM for 2D acceleration at $R=20$.}
    \label{fig:patch_metrics_r20_ssim}
  \end{subfigure}

  \vspace{1ex}  

  \centering
  \begin{subfigure}[t]{0.9\textwidth}
    \centering
    \includegraphics[width=\linewidth]{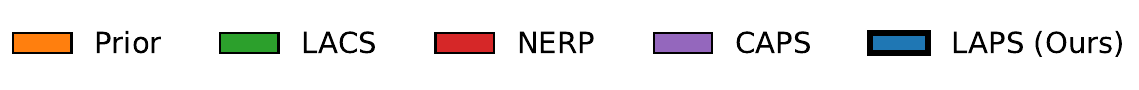}
    \label{fig:sub5}
  \end{subfigure}
  \caption{~Patch-based metric analysis. Each slice in the test set was divided into $32 \times 32$ patches with 50\% overlap, and the similarity between corresponding patches in the target and prior scans was computed using cosine similarity. To evaluate performance as a function of prior similarity, we aggregated all patches across the test set and grouped them into bins based on similarity percentiles. In each subplot, the $x$-axis represents a similarity percentile bin: the leftmost bin includes the 10\% most similar patches, followed by patches in the 10–50\% range (still similar but less so), then the 50–90\% range, and finally, the rightmost bin includes the 10\% most dissimilar patches. The bar labeled Prior shows the metric computed between the target and prior patches within each range.}
\label{fig:patch_metrics}
\end{figure*}

For our patch-based analysis, Fig.~\ref{fig:patch_metrics} compares all longitudinal reconstruction methods and the top-performing non-longitudinal baseline (CAPS) at various similarity percentiles for $R=6$ (1D undersampling) and $R=20$ (2D undersampling). For reference, we include a baseline computed between corresponding patches from the ground truth and prior scans. Results for additional $R$ are provided in Supplementary Fig.~\ref{fig:patch_metrics_sup}.

In the most similar patches (leftmost group), LAPS exceeds CAPS by 1–2 dB in PSNR and 10\% in SSIM, with even larger gains over other longitudinal methods. As similarity decreases, this gap narrows, and in the most dissimilar patches (rightmost group), performance between LAPS and CAPS becomes comparable.

\begin{table*}[t]
\centering
\begin{subtable}{\textwidth}
\centering
\begin{tabular*}{\textwidth}{@{\extracolsep\fill}lcccccc@{\extracolsep\fill}}
\toprule
 & \multicolumn{3}{c}{\textbf{PSNR}} & \multicolumn{3}{c}{\textbf{SSIM}} \\
\cmidrule(lr){2-4} \cmidrule(lr){5-7}
\textbf{Method} & R=3 & R=6 & R=9 & R=3 & R=6 & R=9 \\
\midrule
LAPS (Ours) & \textbf{34.22±3.65} & \textbf{29.32±3.17} & \textbf{27.11±2.94} & 0.925±0.049 & \textbf{0.865±0.069} & \textbf{0.837±0.076} \\
\hline
Target-Prior & 19.93±1.91 & 19.93±1.91 & 19.93±1.91 & 0.677±0.097 & 0.677±0.097 & 0.677±0.097 \\
\hline
AdaDiff & 33.91±3.14 & 26.93±2.44 & 23.26±2.12 & \textbf{0.931±0.045} & 0.832±0.069 & 0.761±0.079 \\
\hline
CAPS & 33.97±3.66 & 28.54±3.10 & 25.46±2.80 & 0.922±0.050 & 0.849±0.072 & 0.802±0.081 \\
\hline
CG-Sense & 30.00±3.57 & 24.63±2.19 & 22.30±1.91 & 0.859±0.075 & 0.778±0.077 & 0.731±0.079 \\
\hline
LACS & 32.61±3.08 & 26.40±2.28 & 23.79±2.04 & 0.920±0.048 & 0.829±0.064 & 0.784±0.069 \\
\hline
MODL & 33.07±4.16 & 27.13±3.43 & 24.17±3.05 & 0.908±0.061 & 0.831±0.077 & 0.786±0.083 \\
\hline
NERP & 31.09±2.71 & 26.31±2.54 & 24.36±2.40 & 0.904±0.055 & 0.822±0.076 & 0.792±0.080 \\
\bottomrule
\end{tabular*}
\caption{1D acceleration}
\label{tab:metrics_1d}
\end{subtable}

\vspace{1em}

\begin{subtable}{\textwidth}
\centering
\begin{tabular*}{\textwidth}{@{\extracolsep\fill}lcccccc@{\extracolsep\fill}}
\toprule
 & \multicolumn{3}{c}{\textbf{PSNR}} & \multicolumn{3}{c}{\textbf{SSIM}} \\
\cmidrule(lr){2-4} \cmidrule(lr){5-7}
\textbf{Method} & R=10 & R=20 & R=30 & R=10 & R=20 & R=30 \\
\midrule
LAPS (Ours) & \textbf{31.12±2.93} & \textbf{28.85±3.44} & \textbf{27.74±3.20} & \textbf{0.875±0.042} & \textbf{0.841±0.054} & \textbf{0.825±0.056} \\
\hline
Target-Prior & 20.11±2.06 & 20.11±2.06 & 20.11±2.06 & 0.688±0.106 & 0.688±0.106 & 0.688±0.106 \\
\hline
AdaDiff & 29.70±3.27 & 25.91±3.13 & 23.84±2.67 & 0.862±0.051 & 0.790±0.066 & 0.746±0.069 \\
\hline
CAPS & 30.48±2.92 & 28.16±3.38 & 26.85±3.13 & 0.861±0.045 & 0.820±0.057 & 0.797±0.059 \\
\hline
CG-Sense & 28.01±2.83 & 25.37±2.71 & 24.06±2.25 & 0.803±0.063 & 0.759±0.068 & 0.732±0.069 \\
\hline
LACS & 29.58±2.65 & 26.62±2.53 & 25.22±2.14 & 0.861±0.045 & 0.810±0.054 & 0.784±0.056 \\
\hline
MODL & 30.15±3.98 & 27.07±4.32 & 25.73±3.91 & 0.855±0.058 & 0.802±0.074 & 0.783±0.073 \\
\hline
NERP & 28.54±2.80 & 26.18±2.97 & 25.20±2.65 & 0.854±0.051 & 0.803±0.063 & 0.783±0.062 \\
\bottomrule
\end{tabular*}
\caption{2D acceleration}
\label{tab:metrics_2d}
\end{subtable}

\caption{~Comparison of PSNR and SSIM for different methods on the test set under (a) 1D acceleration and (b) 2D acceleration at various rates. Best-performing results are highlighted in \textbf{bold}. The row labeled \emph{Target–Prior} reports the metric values computed between the registered target and the prior scans, serving as a baseline for their intrinsic similarity.}
\label{tab:metrics_all}
\end{table*}

\vspace{-1em}\subsection{Performance with Varying Levels of Change}
\begin{figure*}
    \centering
    \includegraphics[width=0.95\linewidth]{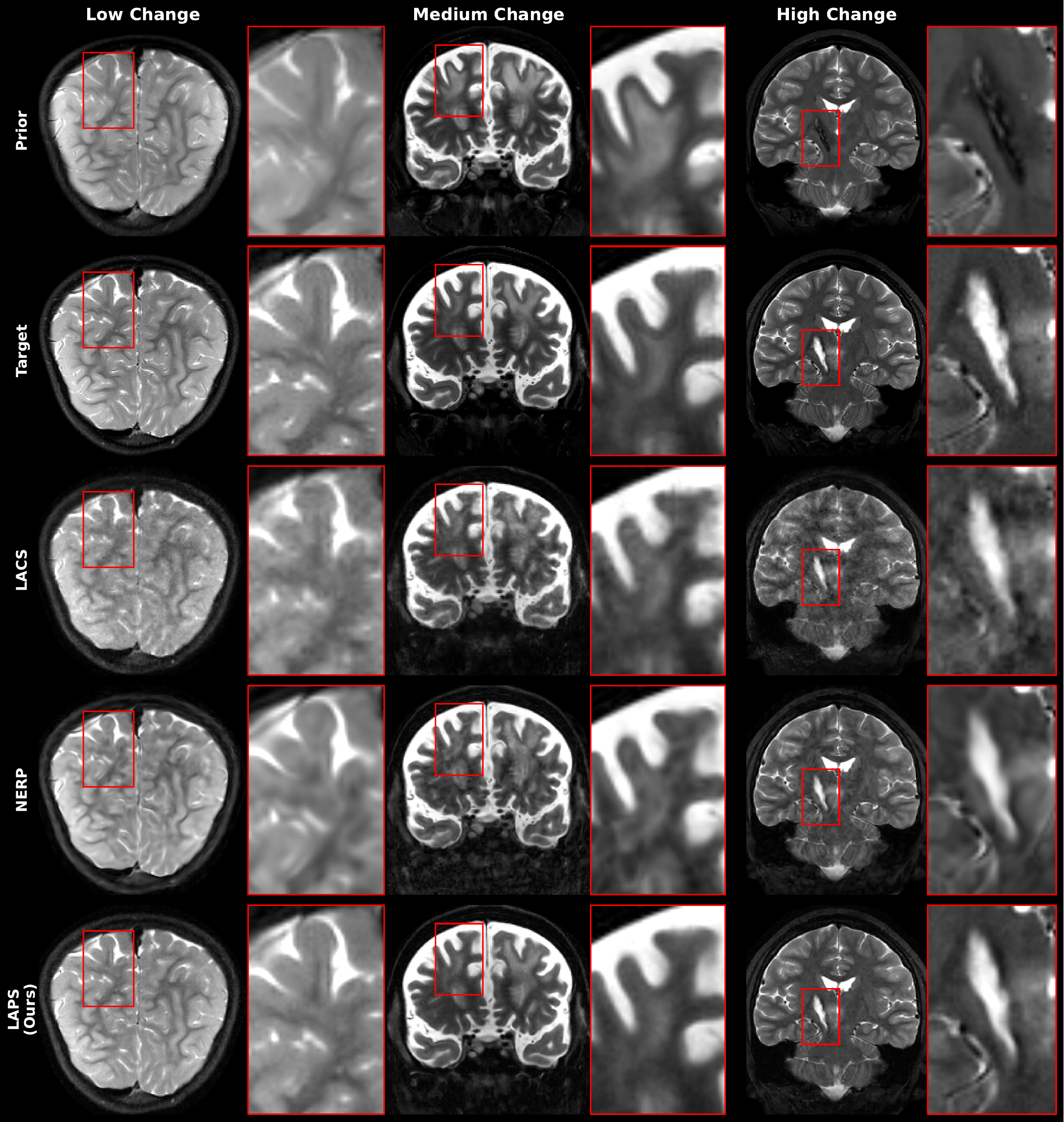}
    \caption{~Comparison of different longitudinal reconstruction algorithms across varying degrees of change between scans, for a 2D undersampling case with $R=20$.
    The first row shows the prior scan, the second row shows the new scan (target), and the subsequent rows present reconstructions from different methods, with our proposed method (LAPS) shown in the last row. Left column: Axial $T_2$-weighted slice from the same patient shown in Fig.~\ref{fig:recon_accel_abnormal}, now displaying a region with minimal or no change between scans.
    Middle column: Coronal $T_2$-weighted images of a patient with a brainstem tumor (not shown). The displayed slice shows $T_2$-hyperintense white matter changes related to treatment, along with white matter swelling that decreases on the follow-up MRI performed 6 days later.
    Right column: Coronal $T_2$-weighted images of a brain with an involuted right basal ganglia hematoma, secondary to a ruptured arteriovenous malformation. The initial exam reveals a $T_2$-hypointense contracted cavity corresponding to old blood products. The follow-up exam, performed 4 months later, shows fluid accumulation within the cavity and $T_2$-hyperintense edema in the surrounding brain parenchyma due to interval radiosurgery treatment. The automatically selected projection times $t_p$, estimated by $\myoperatorname{AutoInit}$, were $280$, $332$, and $453$ for the left, middle, and right columns, respectively, reflecting increasing levels of change across the three cases.}
    \label{fig:diff_abnormal}
\end{figure*}

Fig.~\ref{fig:diff_abnormal} compares longitudinal reconstruction methods across three subjects with increasing levels of change between scan timepoints (left to right), as assessed by a radiologist (see Table~\ref{tab:slam_change}). All reconstructions use 2D undersampling at $R=20$. LACS shows undersampling artifacts in all cases, worsening with greater anatomical change. NERP introduces structural and contrast-related artifacts throughout. In contrast, LAPS maintains consistent reconstruction quality across all levels of change.
Additional high-resolution reconstruction examples with different changes and contrast are provided in Supplementary Figs.~\ref{fig:highres_supp}, \ref{fig:highres_supp_t2f}, and \ref{fig:highres_supp_contrast}.

\vspace{-1em}\subsection{Performance with Misregistration}
\begin{figure*}
    \centering
    \includegraphics[width=0.99\linewidth]{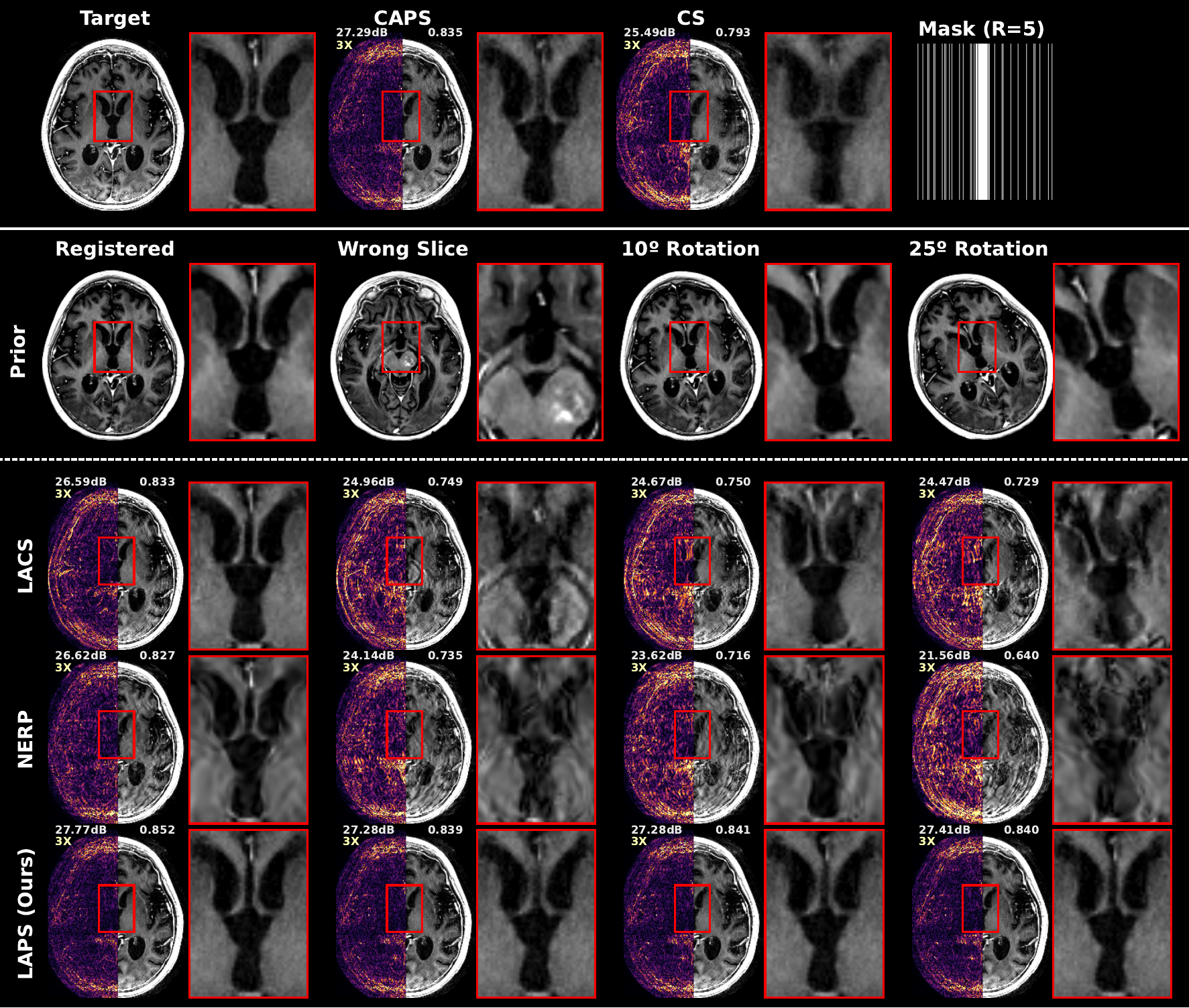}
    \caption{~Reconstruction results of longitudinal methods under various misregistration scenarios, for Axial $T_1$-BRAVO acquisition with $R=5$ 1D undersampling.
    Top row shows the target scan, two non-longitudinal baselines (CAPS and L1-Wavelet CS, which is equivalent to LACS without regularization on the prior image), and the undersampling mask. Below, each column exhibits a different prior registration scenario atop the corresponding longitudinal reconstructions. First is perfect registration, followed by the selection of a prior slice 10mm off from the target slice, finally with mis-alignment of the correct prior slice by $10^\circ$ and $25^\circ$ in-plane rotation offsets, respectively. PSNR and SSIM are shown in the top-left and top-right of each image, respectively. The automatically selected projection times $t_p$, estimated by $\myoperatorname{AutoInit}$ for LAPS were $300$, $450$, $560$, and $600$ (left to right), reflecting decreased levels of reliance on the prior with increased mis-registration.}
    \label{fig:miss_reg}
\end{figure*}

Previous results assumed good alignment between the prior and follow-up scans. To evaluate robustness to misregistration, Fig.~\ref{fig:miss_reg} shows reconstructions of a $T_1$-weighted scan under various misregistration scenarios, including in-plane rotation and slice mismatch, using 1D undersampling at $R = 5$.

Without misregistration, all longitudinal methods perform comparably to non-longitudinal baselines. Under rotation, LACS and NERP exhibit the most severe error, with visible residuals from the misaligned prior. Slice mismatch causes structural interference in both methods, most notably in LACS, which appears as a superposition of the incorrect prior and the follow-up scan. NERP degrades noticeably with increasing misregistration, showing incoherent artifacts. Quantitatively, PSNR drops by 1.5dB or more for LACS and NERP, with NERP decreasing as much as 5dB with the high rotation case. In contrast, LAPS drops by only 0.5dB, maintaining performance comparable to CAPS even under misregistration. Notably, the $\tp$ chosen by $\myoperatorname{AutoInit}$ adapts to the level of misregistration, increasing as misalignment becomes more severe (see figure caption for $\tp$ values).

\vspace{-1em}\section{Discussion}\label{sec:discussion}
\subsection{Effectiveness of Diffusion-Based Priors in Longitudinal Reconstruction}
One of the key factors contributing to the increased performance of LAPS is the use of a powerful LDM trained on large-scale, unpaired data. This enables LAPS to learn a flexible and expressive image prior, in contrast to LACS, which relies on a handcrafted wavelet sparsity prior, and NERP, which uses an INR trained on the prior scan without learning from a broader data distribution. As shown in Tables~\ref{tab:metrics_1d} and \ref{tab:metrics_2d}, LAPS consistently outperforms all other methods across 1D and 2D acceleration settings, maintaining high image quality even at high undersampling rates. The advantage of the LDM prior is further supported by the performance of CAPS, which also utilizes an LDM and shows strong performance compared to the other non-longitudinal methods. Additionally, we found that compared to training an image-space diffusion model like AdaDiff, the LDM training for LAPS took about half as long to converge to meaningful results, while AdaDiff still often struggled to produced reliable reconstructions at high accelerations.

\vspace{-1em}\subsection{Comparison with Prior Longitudinal Methods}
To assess how longitudinal methods handle scan differences, we evaluated performance across varying anatomical change and misregistration. As shown in the patch-based analysis (Fig.~\ref{fig:patch_metrics}), LAPS more effectively leverages regions with high prior similarity. This is also illustrated in Fig.~\ref{fig:recon_accel_normal}, where the prior and follow-up scans differ minimally. In such cases, LAPS selects a smaller $t_p$ (around 200, compared to the typical range of 250-500 observed in the test set), allowing the diffusion process, guided by DC steps, to stay close to the prior while incorporating subtle changes from the measurements.

Importantly, even in regions where the prior offers limited information, LAPS performs on par with or better than CAPS, indicating it does not over-rely on the prior. Instead, it uses high-similarity regions to better condition the global reconstruction, benefiting dissimilar areas as well. No performance degradation is observed in low-similarity patches compared to CAPS, suggesting LAPS effectively balances the LDM and DC, even when the prior is uninformative.

In contrast, LACS’ wavelet sparsity prior becomes insufficient at higher acceleration rates, where the inverse problem is more ill-posed. Furthermore, its prior-weighting scheme depends on intermediate reconstructions, which may be highly degraded under heavy undersampling, limiting its ability to incorporate useful prior information.

NERP performs well when prior and current scans are similar but degrades under larger anatomical changes or severe undersampling (Fig.~\ref{fig:diff_abnormal}, and Supplementary Fig.~\ref{fig:recon_accel_abnormal_full}). In such cases, the INR, which separates use of the prior scan and DC into two different optimization stages, does not have sufficient constraints and often results in overly smoothed or anatomically implausible reconstructions.

Notably, Fig.~\ref{fig:patch_metrics} shows that CAPS performance declines with lower patch similarity, despite it not being a longitudinal method. This is due to the clinical data containing spatially-varying residual noise, causing low cosine similarity in regions with lower SNR rather than true anatomical differences. Consequently, dissimilar patches often have higher noise, reducing metric values for all methods.

\vspace{-1em}\subsection{Robustness to Anatomical Change and Misregistration}
A key requirement for longitudinal reconstruction is preserving meaningful differences between scans. As demonstrated in Figs.~\ref{fig:recon_accel_abnormal} and \ref{fig:diff_abnormal}, LAPS adapts to subtle and large-scale anatomical differences effectively. Its ability to gradually diverge from the prior initialization during reconstruction allows it to capture new pathological structure without forcing agreement with the prior scan.

Additionally, though registering scan timepoints is typically straightforward, it remains important to be robust to misregistration, which can still occur especially when registering to a low-resolution navigator or initial reconstruction. As shown in Fig.~\ref{fig:miss_reg}, LAPS facilitates such flexibility via $\myoperatorname{AutoInit}$, selecting higher $\tp$ with increased prior mis-alignment. In contrast, pixel-domain methods like LACS and NERP exhibit artifacts from misaligned priors. We further attribute LAPS’s robustness to mis-registration to (1) the use of latent representations that are less sensitive to spatial misalignment, and (2) the iterative nature of the diffusion process, which can correct for spatial mismatches by gradually refining the output to match acquired measurements.

\vspace{-1em}\subsection{Limitations and Future Work}
LAPS showcases a feasible approach for improving reconstruction quality with prior scan information, but remains computationally complex. LDM inference requires iterative sampling and repeated DC steps, which is slower than unrolled networks like MODL. However, we note that the LAPS reconstruction time is comparable with AdaDiff and NERP, which both require iterative multi-stage fine-tuning at inference. Future work could explore efficient approximations of the DC update, potentially using learned or linearized operators to reduce runtime.

Additionally, we observed that LAPS, like all methods, experiences degradation at high accelerations, particularly $R\approx30$ in 2D. This performance cap is partially limited by the size and variable quality of SLAM training data, but also from our choice to validate on retrospective clinical cases which undersample k-space suboptimally. To push for higher acceleration, one potential direction is to design sampling patterns tailored for longitudinal imaging. Although previously proposed adaptive k-space sampling methods like those in LACS~\cite{weizman2015compressed} may not yet be practical clinically, an off-line prior-informed strategy for generating  k-space sampling patterns could further improve LAPS at high accelerations.

Though LAPS finds a work-around from needing paired training data, accurately modeling the longitudinal conditional distribution could outperform our inference-time use of the prior. However, our initialization strategy is easy to implement into existing reconstruction strategies, and thus could easily be extended to work synergistically with contrast-to-contrast conditioning or other reconstruction frameworks. One strategy could be to train a generative model conditional on contrast pairs, and use a prior contrast pair to initialize sampling. Such synergy could further refine reconstruction quality, and could even be benchmarked with the multi-contrast SLAM dataset, but likely requires more intensive compute and data given the addition of another representation axis.

\vspace{-1em}\section{Conclusion}\label{sec:conclusion}
The longitudinal application of MRI is vast and is only growing in utility, especially in neuroimaging, both for large-scale research studies as well as numerous clinical settings tracking disease or structural progression \cite{whitwell2008longitudinal}. Consecutive scans of the same subject often contain rich mutual information that, if properly leveraged, can improve reconstruction quality and reduce acquisition time. However, most state-of-the-art reconstruction methods require large, paired datasets for training. This requirement is difficult to satisfy in longitudinal settings, particularly when raw k-space data is needed.

To address this, we proposed LAPS, an LDM-based framework that operates without requiring paired training data. LAPS learns a general image prior from unpaired data and incorporates subject-specific prior scans during inference via a latent-space projection, enabling flexible use of scan priors while preserving clinically meaningful changes. To support evaluation, we also introduce SLAM, a multi-contrast, multi-session clinical dataset with both healthy and pathological cases. SLAM contains raw k-space from follow-up scans, a large portion of which have corresponding prior DICOM scans. The paired subset of SLAM was used here primarily for inference but has potential for supervised training as it grows.

Our experiments show that LAPS consistently outperforms both longitudinal and non-longitudinal baselines across acceleration rates, anatomical changes, and misregistration scenarios. It effectively leverages prior scans when informative and remains robust when they are not. By combining a general LDM prior with subject-specific guidance at inference, LAPS offers a flexible and principled approach to longitudinal reconstruction.

\vspace{-1em}\section*{Acknowledgments}\label{sec:acknowledgements}
\vspace{-1em}
The authors thank Dr. Hugo Decker and Dr. Ali B. Syed for their assistance in the procurement and acquisition of data at the Stanford Hospital that enabled the formation of the SLAM dataset.

\vspace{-1em}\section*{Funding Information}\label{sec:funding_information}
\vspace{-1em}
Grant/Award Number: National Institutes of Health R01 MH116173 \& EB019437.
\vspace{-0.5em}

\bibliography{reference}%

\vspace{-1em}\section*{Supporting Information}
Additional supporting information may be found in the online version of the article at the publisher’s website.

\noindent\textbf{Section S1.} Derivation of the projection timepoint $\tp$, followed by detail on $\myoperatorname{AutoInit}$ and its calibration.

\noindent\textbf{Section S2.} Implementation details for baseline reconstruction algorithms.

\noindent\textbf{Section S3.} Detail on retrospectively under-sampling the already undersampled scans in the SLAM dataset.

\noindent\textbf{Figure S1.} Illustration of optimal $\tp$ selection vs. the prior-target difference as measured by maximization of $\log p (\zgt | \zp)$. Examples of $\tp$ objective function shown for image pairs at two different levels of similarity.

\noindent\textbf{Figure S2.} Calibration development of $\myoperatorname{AutoInit}$ for automatic $\tp$ selection. Relationship between performance for theoretical and empirical $\tp$ optimums plotted as a heatmap, showing strong linear correlation. Linear model fit shown which produces the affine parameters used in Algorithm~\ref{algo:autoinit}. The calibrated $\myoperatorname{AutoInit}$ algorithm is validated against a parameter gridsearch on an unseen test set, outperforming all fixed choices of $\tp$.

\noindent\textbf{Figure S3.} Comparative examples of retrospective undersampling masks. (\subref{fig:mask_retro_1d}) shows 1D undersampling of data acquired with a 1D mask at various rates, as well as a truly fully sampled mask. (\subref{fig:mask_retro_2d}) shows analog 2D cases.

\noindent\textbf{Figure S4.} Comparison of latent posterior sampling inference methods via image quality, quanitative metrics, and reconstruction time. Reconstructions with the same trained network for various inference methods compared: the original PLDS algorithm\cite{rout2023solving}, PLDS initialized at $\tp=200$ with a CG-SENSE recon, our CAPS method, and our proposed LAPS method. LAPS also shown with networks at 4x and 8x down-sampling, highlighting the speed/performance tradeoff for latent compression.

\noindent\textbf{Figure S5.} Performance of reconstruction metrics for various $\nopt$, with $\tp$ chosen via $\myoperatorname{AutoInit}$, for $n_{\rm{step}} = 100$ DDIM steps. $\nopt=1$ equates to sampling with PLDS\cite{rout2023solving}; for both CAPS and LAPS, $\nopt=10$ maximizes PSNR and SSIM in the reconstruction test set. Additionally, LAPS consistently boosts performance compared to CAPS for each value of $\nopt$, showing the utility of the longitudinal prior.

\noindent\textbf{Figure S6.} Expansion of Fig.~\ref{fig:recon_accel_normal}, shown for all methods.

\noindent\textbf{Figure S7.} Patch based histogram similar to the ones shown in Fig.~\ref{fig:patch_metrics}, for $R=3,5$ and $9$ (1D), and $R=10, 15$ and $30$ (2D).

\noindent\textbf{Figure S8.} Additional reconstruction example for all methods at R=7 with 1D undersampling, shown as an axial view for the large change coronal example in Fig.~\ref{fig:diff_abnormal}.

\noindent\textbf{Figure S9.} Additional reconstruction example for all methods at R=20 with 2D undersampling for a $T_2$-FLAIR contrast.

\noindent\textbf{Figure S10.} Additional reconstruction example for all methods at R=7 with 1D undersampling, for an example where the prior contrast slightly differs from the new scan contrast.

\clearpage
\setcounter{page}{1}
\addcontentsline{toc}{section}{Supplementary Information}

\setcounter{section}{0}
\renewcommand{\thesection}{S\arabic{section}}

\renewcommand{\thesubsection}{S\arabic{section}.\arabic{subsection}}

\setcounter{figure}{0}
\renewcommand{\thefigure}{S\arabic{figure}}
\setcounter{table}{0}
\renewcommand{\thetable}{S\arabic{table}}
\setcounter{equation}{0}
\renewcommand{\theequation}{S\arabic{equation}}

\vspace{-1em}\section{Projection Timepoint $\tp$}\label{sec:projection_timepoint}

\vspace{-0.5em}

\subsection{Derivation for Projection Timepoint $\tp$ in the Gaussian Case}\label{sec:tp_apdx}

\vspace{-0.75em}

Here we compute the optimal projection point $t_p$ for the case $\bz_0 \sim \cN(0, \bI)$, as the typical training objective of VAEs minimizes divergence of the learned $p(\bz)$ from a normal Gaussian\cite{kingma2013auto}. Beginning with Eq.~\eqref{eq:opt_tp}:
\begin{align}
t_p &= \argmax{t}{p_0(\bz_0 = \zgt | \bz_t = \zp_t)} \\
&= \argmax{t}{\frac{p_0(\bz_0 = \zgt | \bz_t = \zp_t)}{p_0(\bz_0=\zgt)}} \label{eq:tp_1}\\
&= \argmax{t}{\frac{p_t(\bz_t = \zp_t | \bz_0 = \zgt)}{p_t(\bz_t = \zp_t )}} \label{eq:tp_2}\\
&= \argmax{t}{\frac{p_t(\bz_t = \sqrt{\balph}\zp + \sqrt{1 - \balph}\beps | \bz_0 = \zgt)}{p_t(\bz_t = \sqrt{\balph}\zp + \sqrt{1 - \balph}\beps)}}\label{eq:tp_3}\\
&= \argmax{t}{\frac{\cN(\sqrt{\balph}\zp ; \sqrt{\balph}\zgt, 2 (1-\balph))}{\cN(\sqrt{\balph}\zp ; 0, 2 - \balph)}}\label{eq:tp_4}\\
&= \argmax{t}{\frac{\sqrt{2-\balph}}{\sqrt{2(1-\balph)}}\cdot \frac{\exp\{-\frac{\balph}{4(1-\balph)}\|\zgt - \zp\|_2^2\}}{\exp\{-\frac{\balph}{2(2-\balph)}\|\zp\|_2^2\}}}\\
&= \argmax{t}{\gamma_t +\frac{\balph}{2}\left[\frac{\|\zp\|_2^2}{2-\balph} - \frac{\|\zgt - \zp\|_2^2}{2(1-\balph)}\right]} \label{eq:tp_math_sol}
\end{align}
Where,
\begin{equation}
    \gamma_t = \frac{1}{2}\log(\frac{2-\balph}{2(1-\balph)})
\end{equation}
The derivation steps follow as: Step~\eqref{eq:tp_1} divides by a term independent of $t$; Step~\eqref{eq:tp_2} applies Bayes’ rule; Step~\eqref{eq:tp_3} substitutes the definition of $\bz_t$ from Eq.~\eqref{eq:proj_zp}; and Step~\eqref{eq:tp_4} uses the forward diffusion process, with $\beps \sim \cN(0,\bI)$ independent of $\bz_0 \sim \cN(0,\bI)$.  

\begin{figure*}[bh]
    \centering
    \includegraphics[width=0.9\linewidth]{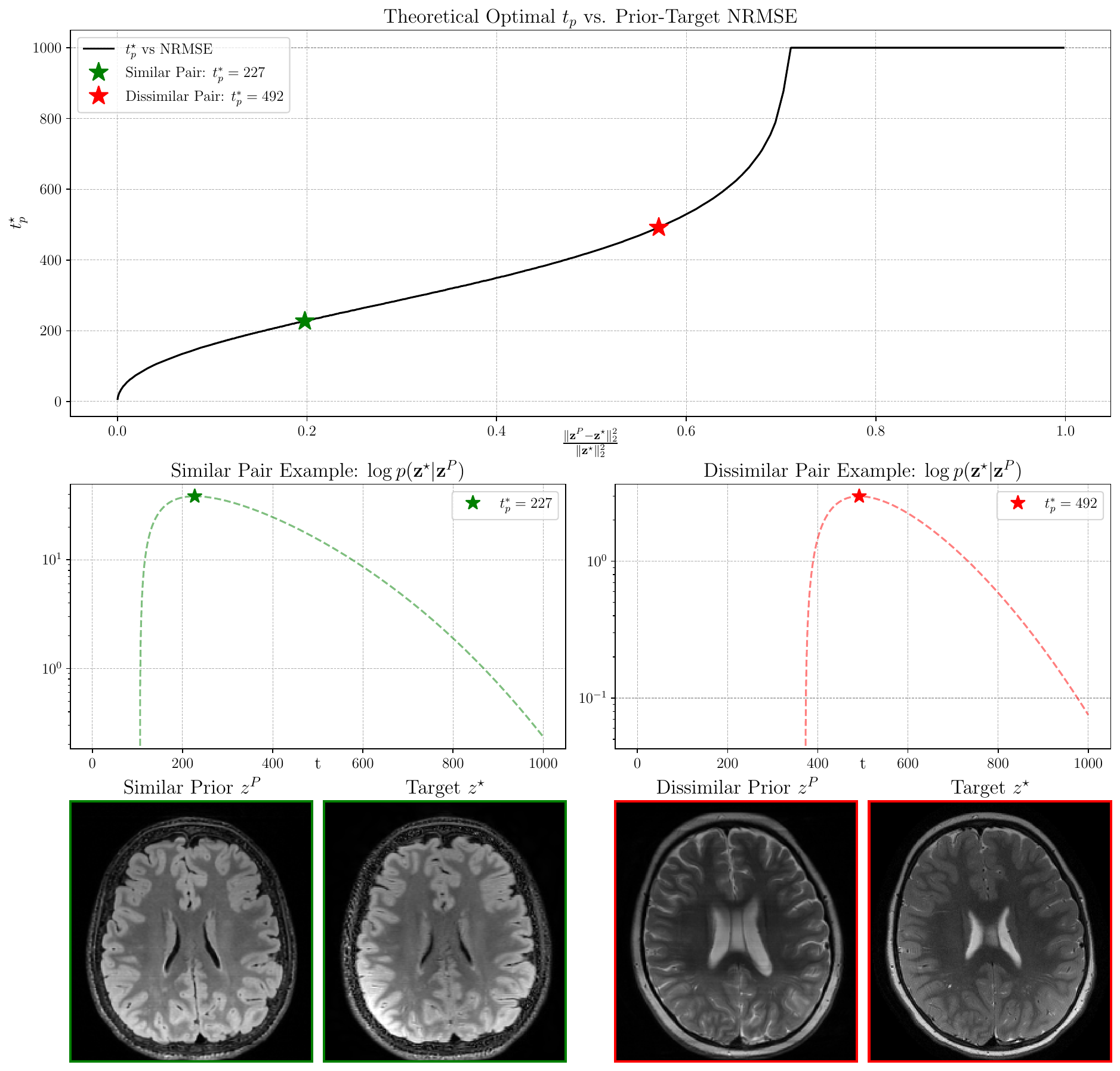}
    \caption{~Illustration of optimal $t_p$ selection vs. prior-target difference. Top sub-plot shows the optimal $\tp$ when $\zgt$ is selected randomly, and $\zp = \zgt +\bn$, over different levels of additive noise and thus different prior-target NRMSE. Below, examples of the objective in Eq.~\eqref{eq:tp_math_sol} which proxies $\log p_0(\zgt \mid \bz_t = \zp_t)$ are shown for real image pairs, with the respective curve maxima as the selected $\tp^\star$'s. Shown on the left (green) is a pair where the target and prior are highly similar (low NRMSE), and on the right (red) is a pair with increased dis-similarity (higher prior/target NRMSE), with visible difference in contrast and morphology in the central ventricles. As can be seen, the optimal $\tp$ for our learned latent space is some intermediate timestep for both cases, with higher image dis-similarity favoring higher $\tp^\star$.}
    \label{fig:tp_sim}
\end{figure*}

To illustrate the behavior of Eq.~\eqref{eq:tp_math_sol} as a function of scan pair similarity, we sample $\zgt \sim \cN(0, \bI)$ and define $\zp = \zgt + \bn$, with $\bn$ a random noise vector. Shown in Fig.~\ref{fig:tp_sim}, we compute the optimal $t_p$ as a function of latent NRMSE by varying the noise level:
\begin{equation}
    \myoperatorname{NRMSE}(\zp, \zgt) = \frac{\|\zp-\zgt\|_2^2}{\|\zgt\|_2^2},
\end{equation}
Here, lower NRMSE corresponds to smaller optimal $t_p$, while higher NRMSE yields larger $t_p$, reflecting reduced reliance on the prior when similarity is poor. Fig.~\ref{fig:tp_sim} additionally shows that Eq.~\eqref{eq:tp_math_sol} is concave for specific scan pairs, shown at two different similarity levels.
\vspace{-1em}

\vspace{-0.5em}

\subsection{Details of $\myoperatorname{AutoInit}$}\label{sec:auto_init_calib}

\vspace{-0.5em}

We address both challenges of prior initialization using an initial reconstruction $\xinit$. 
To overcome the lack of phase in $\xp$, we extract $\phi = \angle \xinit$ and use it to initialize the projection. 
To adapt $\tp$ to scan variability, we use $\zinit = \cE(\xinit)$ as a proxy for $\zgt$. 
Assuming $\|\zinit - \zgt\|_2$ is reasonably small, Eq.~\eqref{eq:timeproject} can be approximated as
\begin{equation}\label{eq:tp_approx_appendix}
\tpapprox = \myoperatorname{TimeProject}(\zp, \zinit).
\end{equation}

To validate this approximation and other assumptions in deriving $\myoperatorname{TimeProject}$, we chose to empirically calibrate $\tpapprox$ using a small paired validation dataset $\Validation = \{(\xgt_i, \xp_i, \by_i)\}_{i=1}^{N_v}$. 
For each validation example, we run reconstructions across a grid of $\tp$ values from $T$ down to $0$, additionally computing $\tpapprox$ from $\xinit$. 
The empirically optimal projection $\tpemp$ is identified for each case based on reconstruction metrics. 
As shown in Fig.~\ref{fig:auto_tp_c}, the relation between $\tpemp$ and $\tpapprox$ is approximately linear. 
We therefore define our calibration as an affine mapping of $\tpapprox$ to $\tpemp$, yielding scale and shift parameters $v_p$ and $w_p$ used to get the final $\tp$ in Algorithm~\ref{algo:autoinit} such that $\tp = \tpscale \cdot \tpapprox + \tpshift \approx \tpemp
$.

\vspace{-0.75em}\subsubsection{Implementation Details of Calibration}\label{sec:auto_init_calib_implementation}
\vspace{-0.5em}

Calibration was performed using 100 slices sampled across 10 subjects and undersampling strategies (i.e. forward models) in the SLAM training set, which we found sufficient to provide reliable $t_p$ values. This sufficiency arises from the fact that the relationship between $\tpapprox$ and $\tpemp$ is well captured by an affine transform. 
Reconstructions across the $t_p$ grid were obtained using our accelerated CAPS network as $\myoperatorname{InitReconAlg}$ to generate $\xinit$, with reconstruction metrics binned by the computed $\tpapprox$ (shown as a heatmap in Fig.~\ref{fig:auto_tp_a}) to identify the trend in empirical optimal $\tpemp$ over scan pair differences. 
The fitted affine parameters were $\tpscale=1.55$ and $\tpshift=-350$, shown as slope and intercept of the trend in Fig.~\ref{fig:auto_tp_c}. We verified that these generalized to unseen test subjects in Figs.~\ref{fig:auto_tp_d},~\ref{fig:auto_tp_e}, outperforming any fixed $\tp$ choice. We henceforth use this affine-adjusted projection at inference for all experiments. 

$\myoperatorname{InitReconAlg}$ can in principle be any sufficiently fast method that provides a stable approximation of the target image; here we used CAPS due to its speed and robustness under aggressive acceleration. Fig.~\ref{fig:auto_tp_b} additionally shows our choice in $\zinit$ well proxies $\zgt$ for $\myoperatorname{TimeProject}$, though introducing additional variance. 

\vspace{-0.75em}\subsubsection{Paired Dataset Requirements}
\begin{figure*}[ht]
  \begin{center}
    \includegraphics[width=\linewidth]{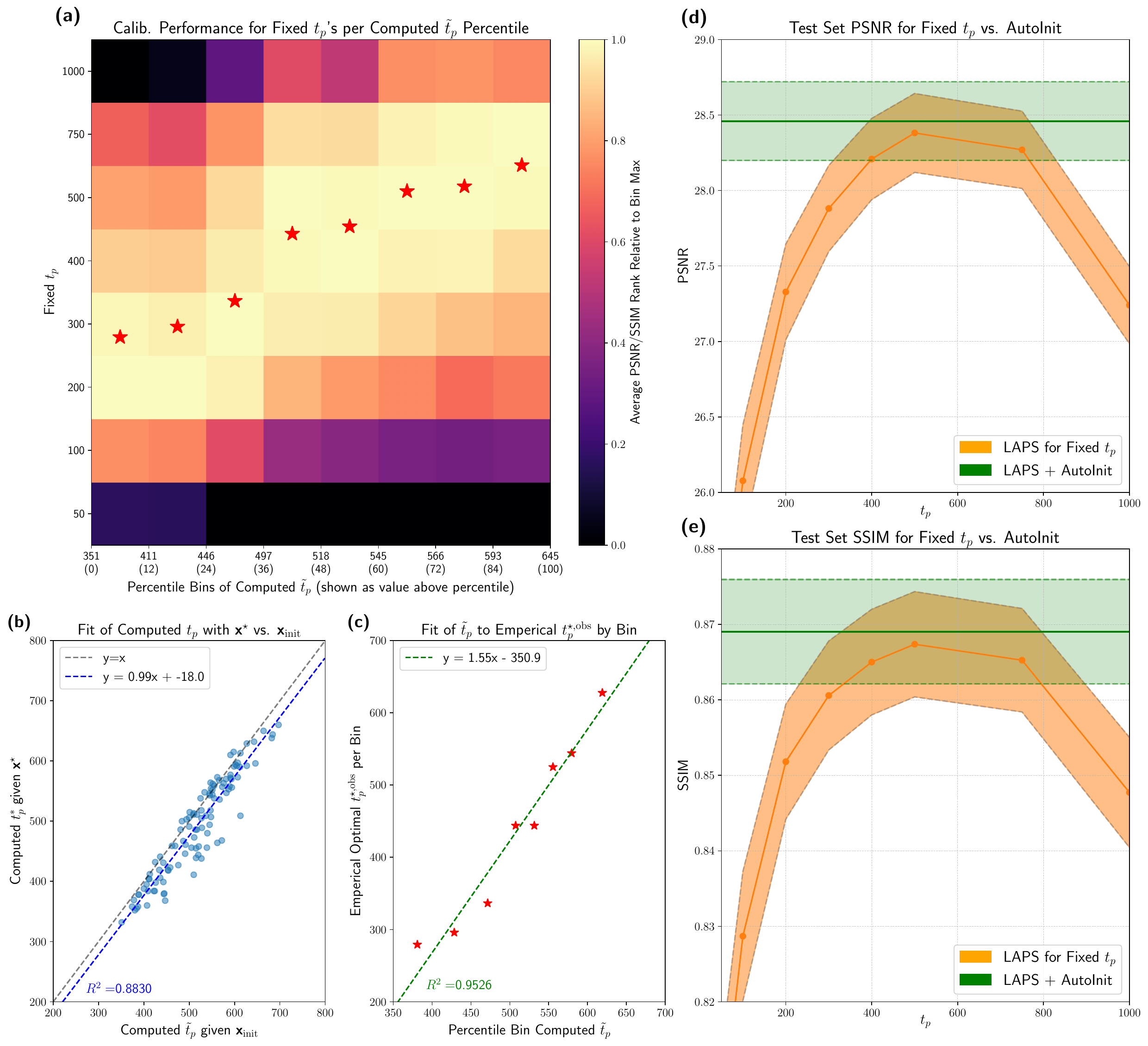}

    \begin{minipage}[t]{0pt}\phantomsubcaption\label{fig:auto_tp_a}\end{minipage}
    \begin{minipage}[t]{0pt}\phantomsubcaption\label{fig:auto_tp_b}\end{minipage}
    \begin{minipage}[t]{0pt}\phantomsubcaption\label{fig:auto_tp_c}\end{minipage}
    \begin{minipage}[t]{0pt}\phantomsubcaption\label{fig:auto_tp_d}\end{minipage}
    \begin{minipage}[t]{0pt}\phantomsubcaption\label{fig:auto_tp_e}\end{minipage}
    \vspace{-2.25em}
    \caption{~Development of $\myoperatorname{AutoInit}$ for Automatic $\tp$ Selection. To empirically validate and calibrate $\myoperatorname{AutoInit}$, we run LAPS at several fixed values of $\tp$ at R=7 with 1D undersampling and R=25 for 2D undersampling for 100 slices in the calibration set $\Validation$, which is comprised of a mix of registered and un-registered prior-target pairs. Per pair, we compute $\tpapprox$ via Eq.~\eqref{eq:tp_approx} using a fast CAPS recon for $\xinit$. We validate that Eq.~\eqref{eq:tp_approx} is a good approximation for  Eq.~\eqref{eq:opt_t_ex} in (\subref{fig:auto_tp_b}), where the $\tp$ computed given the ground truth $\xgt$ and prior $\xp$ correlates to our computed $\tilde{t}_p$ given $\xinit$ and $\xp$ with $R^2=0.883$ and trendline slope of nearly 1. We then split $\Validation$ into 8 bins by the percentile of the computed $\tpapprox$. Per bin, we compute the average PSNR and SSIM of the reconstructions for each fixed $\tp$, assigning a score as the average of the rank in both metrics over all fixed $\tp$'s. This score is normalized per bin such that the lowest and highest performing fixed $\tp$'s score 0 and 1, respectively. (\subref{fig:auto_tp_a}) plots a heatmap of these relative scores across all bins and fixed $\tp$'s. We additionally denote the location of the empirical optimal $\tpemp$ per bin with a red star, found as the argmax over a spline-smooth curve fit to each column of the heatmap. (\subref{fig:auto_tp_c}) shows a linear regression fit of the per-bin $\tpemp$ values against the computed $\tpapprox$ at the center of each bin, finding $\tpscale=1.55, \tpshift=-350$ from the regression slope and intercept as the $\myoperatorname{AutoInit}$ calibration parameters . Finally, we selected a random subset of 70 slices in the test set comprised of registered and mis-registered prior-target pairs, and run LAPS at the same fixed $\tp$ values, in addition to a LAPS reconstruction using the newly affine-calibrated $\myoperatorname{AutoInit}$ estimation of $\tp$. (\subref{fig:auto_tp_d}) and (\subref{fig:auto_tp_e}) show the PSNR and SSIM, respectively, across reconstructions with the different choices of $\tp$, shown as the mean metric value over the dataset $\pm$ standard error. As $\tp$ increases, initially both metrics increase, but with too high of $\tp$, performance decreases. Note that $\tp=1000=T$ is equivalent to unconditional inference, as the initialization is pure noise. However, $\myoperatorname{AutoInit}$ outperforms all choices of fixed $\tp$, demonstrating the flexibility of this prior initialization.}
    \label{fig:auto_tp}
  \end{center}
\end{figure*}

\vspace{-0.5em}
Although SLAM includes longitudinal pairs, our method does not rely on paired data for LDM training. We only use a small calibration set described above to tune $\myoperatorname{AutoInit}$, but this does not require modeling the conditional distribution between $\zp$ and $\zgt$ (which would demand more paired data than SLAM currently provides). Instead, the calibration simply relates latent distances to the unconditional diffusion process. Without such calibration, our analysis of performance vs. $\tp$ in Figs.~\ref{fig:auto_tp_d},~\ref{fig:auto_tp_e} shows that using a fixed $\tp = 450$ still performs comparably to our adaptive $\tp$ solution, though with slightly reduced overall performance.

\clearpage
\onecolumn
\setcounter{page}{1}
\vspace{-1em}\section{Baseline Implementation Details}\label{sec:baseline_apdx}
Below we summarize the hyperparameters and implementation details for each baseline method.

\begin{enumerate}[leftmargin=*]
    \item \textbf{LACS:}  
    We solve the weighted $\ell_1$-regularized problem combining wavelet sparsity and similarity to a prior scan:    
    \begin{equation}
        \argmin{\bx}{\|\bcA\bx - \by\|_2^2
        	+ \lambda_1 \|\bW_1 \bPsi \bx\|_1
        	+ \lambda_2 \|\bW_2(\bx - \xp)\|_1.}
    \end{equation}
    Here $\bPsi$ is a wavelet transform, and $\bW_1, \bW_2$ are spatially adaptive weights updated iteratively. 
    Hyperparameters were tuned via validation, yielding $\lambda_1 = 10^{-5}$ and $\lambda_2 = 10^{-4}$. 
    For fairness, we omit the adaptive acquisition component originally proposed in the original paper~\cite{weizman2015compressed}.

    \item\textbf{NERP:}  
    We represent the prior $\xp$ using an INR parameterized by an MLP mapping spatial coordinates $(x,y)$ to the complex image at that location. 
    As the original work used magnitude-only simulated data, we adapt NERP to complex data by estimating the initial phase from a CG-SENSE reconstruction before pre-training (similar to how LAPS uses a fast CAPS initialization).  
    After this initialization, the INR is fine-tuned through the measurement model. We use the same hyperparameters and network structure as detailed in the original paper~\cite{shen2022nerp}.
    
    \item \textbf{MODL.}  
    We follow the setup in~\cite{aggarwal2018modl}, unrolling conjugate gradient (CG) updates with CNN-based denoising blocks, trained end-to-end. 
    Since MODL requires the forward model during training, we trained separate models for each acceleration rate, with all other hyperparameters matching the original paper. As training requires paired k-space and reconstructions, we train MODL only on the SLAM partition of follow-up scans per acceleration rate. Each model was trained with 10 unroll iterations for 1 day on a single A6000 GPU.

    \item \textbf{AdaDiff.}  
    We use the official open-source implementation of AdaDiff~\cite{dar2022adaptive}, training on the same dataset as LAPS. 
    The method employs a two-stage framework: (1) fast diffusion-based reconstruction with a small number of denoising steps, augmented by data consistency operations; (2) fine-tuning of the $t=0$ denoiser using the acquired measurements to improve fidelity. 
    We selected AdaDiff as it provides strong performance among recent image-space diffusion models, serving as a representative benchmark against state-of-the-art diffusion-based reconstructions. AdaDiff was trained for 14 days across 5 A6000 GPUs, which more than doubles the training time used for the LDM in our LAPS framework.
    
    \item \textbf{CAPS.}  
    We use the same hyperparameters as LAPS, except we fix $\tp$ for initialization. This fixed $\tp$ was used as the distribution shift between the initializing CG reconstruction projected to $\tp$ and the optimal reconstruction latent $\zgt$ remains relatively constant. A hyperparameter search showed $\tp=200$ to perform best for CAPS.
    All other details follow Section~\ref{sec:latent_recon}.
\end{enumerate}

\clearpage
\setcounter{page}{1}
\vspace{-1em}\section{Further Undersampling Already Undersampled Scans}\label{sec:undersampling}

\begin{figure*}
  \begin{center}
    \begin{subfigure}[b]{0.95\textwidth}
      \centering
      \includegraphics[width=\linewidth]{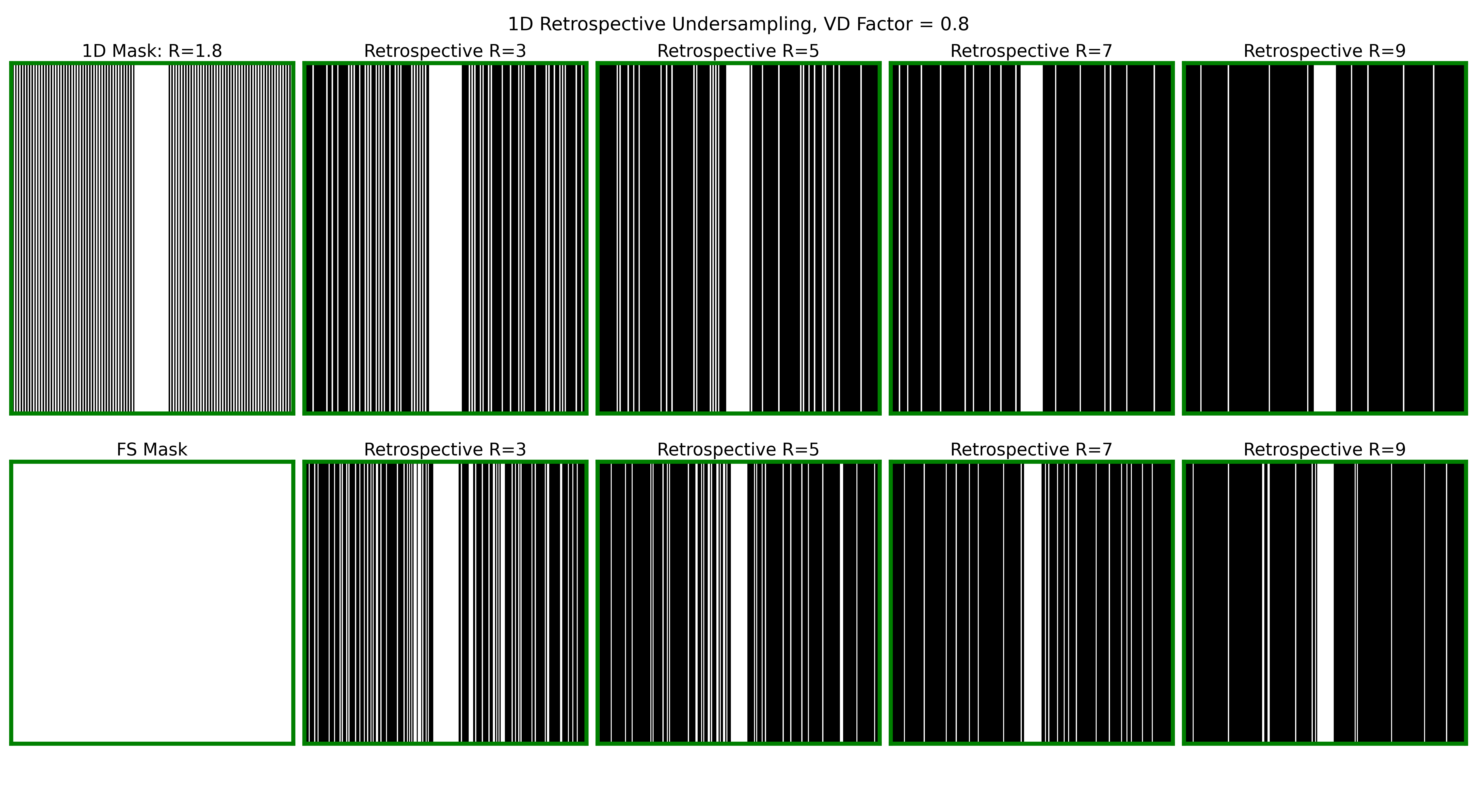}
      \vspace{-3em}
      \caption{Example Retrospective Undersampling with 1D mask from 1D undersampled scan compared to fully sampled scan.}
      \label{fig:mask_retro_1d}
    \end{subfigure}%
    \vspace{1ex}  
    \begin{subfigure}[b]{0.95\textwidth}
      \centering
      \includegraphics[width=\linewidth]{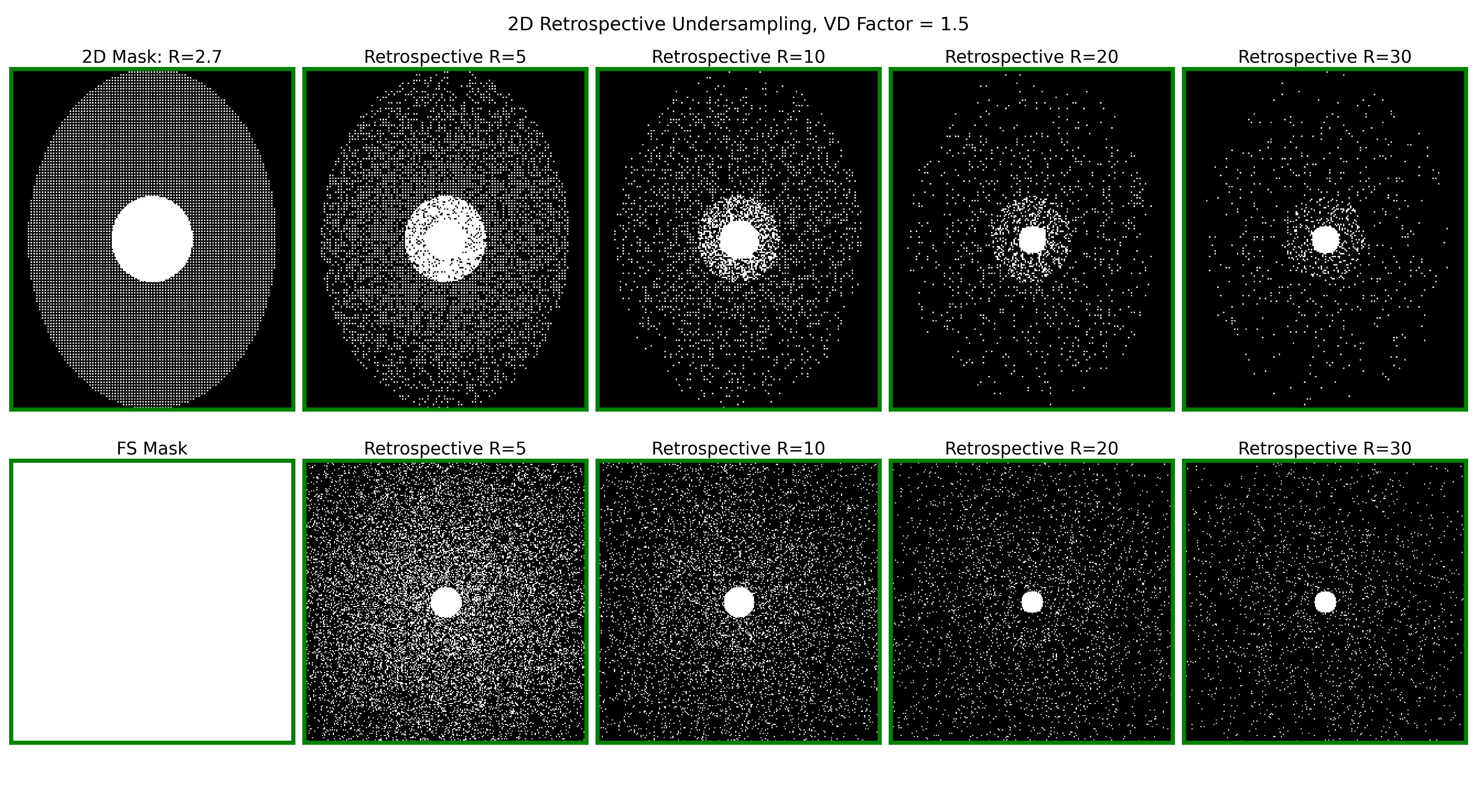}
      \vspace{-3em}
      \caption{Example Retrospective Undersampling with 2D mask from 2D undersampled scan compared to fully sampled scan.}
      \label{fig:mask_retro_2d}
    \end{subfigure}%
  \caption{~Comparison of retrospective undersampling masks. (\subref{fig:mask_retro_1d}) shows 1D undersampling of data acquired with a 1D mask (top) at various rates, as well as a truly fully sampled mask (bottom). (\subref{fig:mask_retro_2d}) shows the analog 2D cases.}
  \vspace{-1ex}  
  \label{fig:retrospective_undersampling}
  \end{center}
\end{figure*}

To further undersample raw k-space retrospectively from our SLAM dataset which may have already been undersampled, we implement the following procedure:

\begin{enumerate}
    \item If a scan was collected as a 2D acquisition with in-plane acceleration, or as a 3D acquisition with only one undersampling direction, the scan can be reduced to a set of slices each collected with the same 1D undersampling pattern. In this case, to futher retrospectively undersample each slice, we first form a 1D undersampling mask selecting for all $N_{\rm{samp}}$ acquired phase encoding lines less than or equal to the phase encoding matrix dimension $N$, as in the top-left subplot of Fig.~\ref{fig:mask_retro_1d}. In our dataset, these masks always have a fully sampled center region of width $N_{\rm{fs, full}} < N$ sufficient for sensitivity map estimation. To further undersample this data to rate $R$, we first set a fully sampled center width $N_{\rm{fs}}$ to retain with the following heuristic:
\begin{equation}
    N_{\rm{fs}} = \begin{cases}
        \min(N_{\rm{fs, full}}, 21) \quad R \leq 3 \\
        \min(N_{\rm{fs, full}}, 15) \quad R > 3 \\
    \end{cases}
\end{equation}
    Denoting the set of sampled lines outside of the retained fully sampled region as $\mathfrak{N}$ with size $|\mathfrak{N}| = N_{\rm{samp}} - N_{\rm{fs}}$, we sample $N_{\rm{ext}}=\lfloor \frac{N}{R} \rfloor -N_{\rm{fs}}$ exterior lines from $\mathfrak{N}$ without replacement via a multinomial distribution, where each exterior line $i$ with a distance of $d_i$ points from k-space center is sampled with probability:
 \begin{equation}
    p_i =\frac{
      \left (1 - \frac{\gamma_{\rm{1D}}}{N/2} d_i \right )^{\alpha_{\rm{VD}}}
    }{
      \sum_{j=1}^{|\mathfrak{N}|} \left (1 - \frac{\gamma_{\rm{1D}}}{N/2} d_j \right )^{\alpha_{\rm{VD}}}
    }
\end{equation}   
    where $\alpha_{\rm{VD}}$ is a density factor that shifts probability density mass to the center of k-space with increasing values. In our experiments, we use $\alpha_{\rm{VD}}=0.8, \gamma_{\rm{1D}}=0.9$ for 1D undersampling cases. 
    
    Finally, because random sampling with variable density can leave large gaps in k-space, instead of sampling these $N_{\rm{ext}}$ exterior lines once, to form each mask we instead sample the set of exterior lines 100 times, and use the set in which the maximum distance between sampled lines is minimized. The final retrospective mask then contains these $N_{\rm{ext}}$ variable-density sampled exterior lines with the $N_{\rm{fs}}$ retained fully sampled lines, shown for various rates in the right columns of Fig.~\ref{fig:mask_retro_1d}. 

    \item If a scan was collected in 3D with two undersampled dimensions, as in the top-left subplot of Fig.~\ref{fig:mask_retro_2d}, we first iFFT the acquired k-space along the fully-sampled readout dimension and form 2D slices in which the undersampling dimensions become in-plane. Because these in-plane masks are sometimes anisotropic or non-rectangular, we then designate the number of fully-sampled points in the acquisition as the number of points in the tightest bounding geometry over the acquired mask (in the case of Fig.~\ref{fig:mask_retro_2d}, this would be an elliptical mask). We denote the total points in this bounded region as $N$ and the sampled points in the mask as $N_{\rm{samp}}$. Our dataset's 2D undersampled examples always exhibit some fully-sampled center region which supersets a circular region of diameter $N_{\rm{fs, 2D, full}}$ points. To further undersample this bounded region to rate $R$, we first set a fully sampled center diameter $N_{\rm{fs, 2D}}$ to retain a circular center region for sensitivity map estimation with the following heuristic:
\begin{equation}
    N_{\rm{fs, 2D}} = \begin{cases}
        \min(N_{\rm{fs, 2D, full}}, 25) \quad R \leq 15 \\
        \min(N_{\rm{fs, 2D, full}}, 14) \quad R > 15 \\
    \end{cases}
\end{equation} 
    Denoting the set of sampled k-space points outside of this retained fully-sampled circle as $\mathfrak{N}_{\rm{2D}}$, we now sample $N_{\rm{ext, 2D}} = \lfloor\frac{N}{R} \rfloor - \lfloor \pi (\frac{N_{\rm{fs, 2D}}}{2})^2 \rfloor$ points, each with probability:
 \begin{equation}
    p_i =\frac{
      \left (1 - \gamma_{\rm{2D}} d_i \right )^{\alpha_{\rm{VD,2D}}}
    }{
      \sum_{j=1}^{|\mathfrak{N}_{\rm{2D}}|} \left (1 - \gamma_{\rm{2D}} d_j \right )^{\alpha_{\rm{VD,2D}}}
    }
\end{equation}   
    where each sampleable exterior point has a distance $d_i$ from the center, normalized such that the furthest point has a distance of 1. We use $\alpha_{\rm{VD, 2D}}=1.5, \gamma_{\rm{2D}}=0.8$. In the 2D case, we only sample exterior points once for simplicity. The final mask combines the exterior sampled points with the retained fully sampled region.
    
    \item If a scan was collected without acceleration (i.e. fully sampled), we retrospectively can undersample it both with 1D undersampling and 2D undersampling, as described in the above procedures. Examples of this at various rates are shown for the bottom rows of Fig.~\ref{fig:mask_retro_1d} and Fig.~\ref{fig:mask_retro_2d}.
\end{enumerate}

\clearpage
\setcounter{page}{1}
\onecolumn
\vspace{-1em}\section{Supplementary Figures}\label{sec:supp}

\begin{figure}[bh]
  \begin{center}
    \begin{subfigure}[t]{\textwidth}
      \centering
      \includegraphics[width=\linewidth]{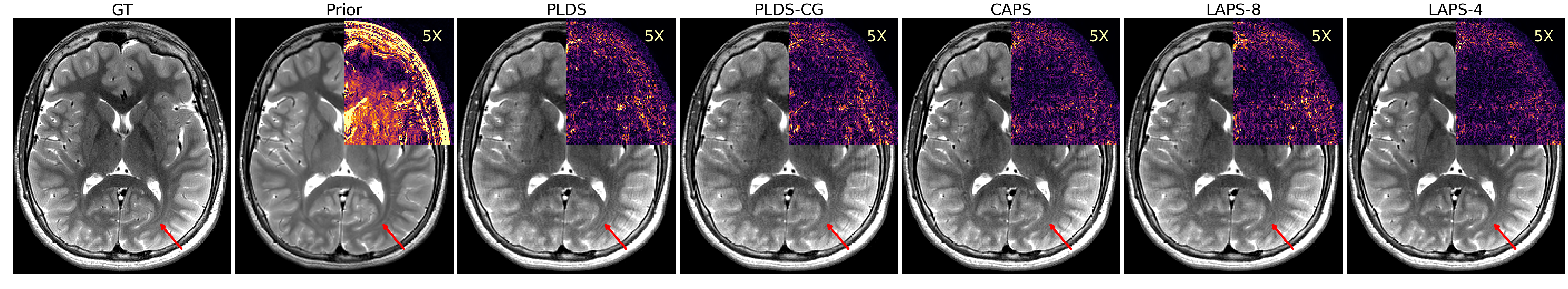}
      \caption{Example Reconstruction with various latent posterior sampling methods.}
      \label{fig:latent_sampling_upper}
    \end{subfigure}%
    \vspace{1ex}  
    \begin{subfigure}[t]{\textwidth}
      \centering
      \includegraphics[width=\linewidth]{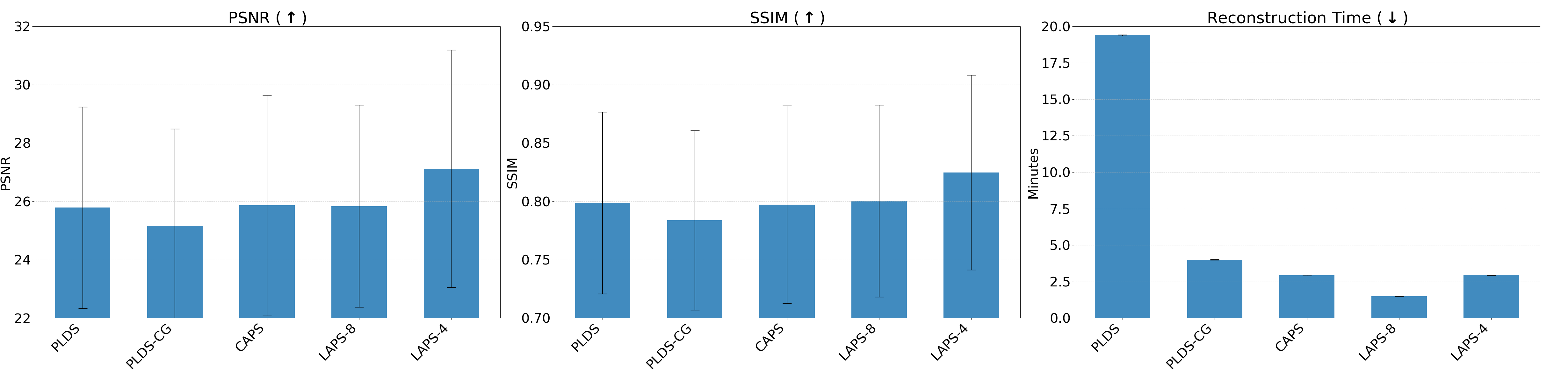}
      \caption{Cumulative Metrics over a subset of 40 slices in the test set, with a mix of R=25 acceleration with 2D undersampling, and R=7 acceleration with 1D undersampling.}
      \label{fig:latent_sampling_lower}
    \end{subfigure}%
  \caption{~Comparison of Latent Posterior Sampling Inference Methods. All methods use the same trained networks with VAE downsampling of $K=4$, with the exception of LAPS-8, which uses equivalent networks with further downsampling to $K=8$. (\subref{fig:latent_sampling_upper}) shows a typical reconstruction example with the upper-right quartile of the image replaced with 5x error with respect to the ground truth (GT). From left to right: PLDS\cite{rout2023solving}, an extension of LDPS which uses a DDPM sampler for one latent gradient step per DC update, requires substantially longer reconstruction times, and undersampling artifacts are still visible. PLDS-CG is an initialization of PLDS at $\tp=200$ with a CG-SENSE recon, decreasing reconstruction time 5x as shown in (\subref{fig:latent_sampling_lower}), but with a minor performance hit. CAPS, our proposed latent sampling method without a prior scan, is also initialized at $\tp=200$ with the CG-SENSE recon, but instead uses a DDIM sampler with repeated DC for $\nopt = 10$, speeding up reconstruction further with equivalent performance to PLDS. Finally, LAPS-8 and LAPS-4 show our method at different down-sampling levels, highlighting the speed/performance tradeoff with latent compression.}
  \label{fig:latent_sampling_methods}
  \end{center}
\end{figure}

\begin{figure}[bh]
    \centering
    \includegraphics[width=0.90\linewidth]{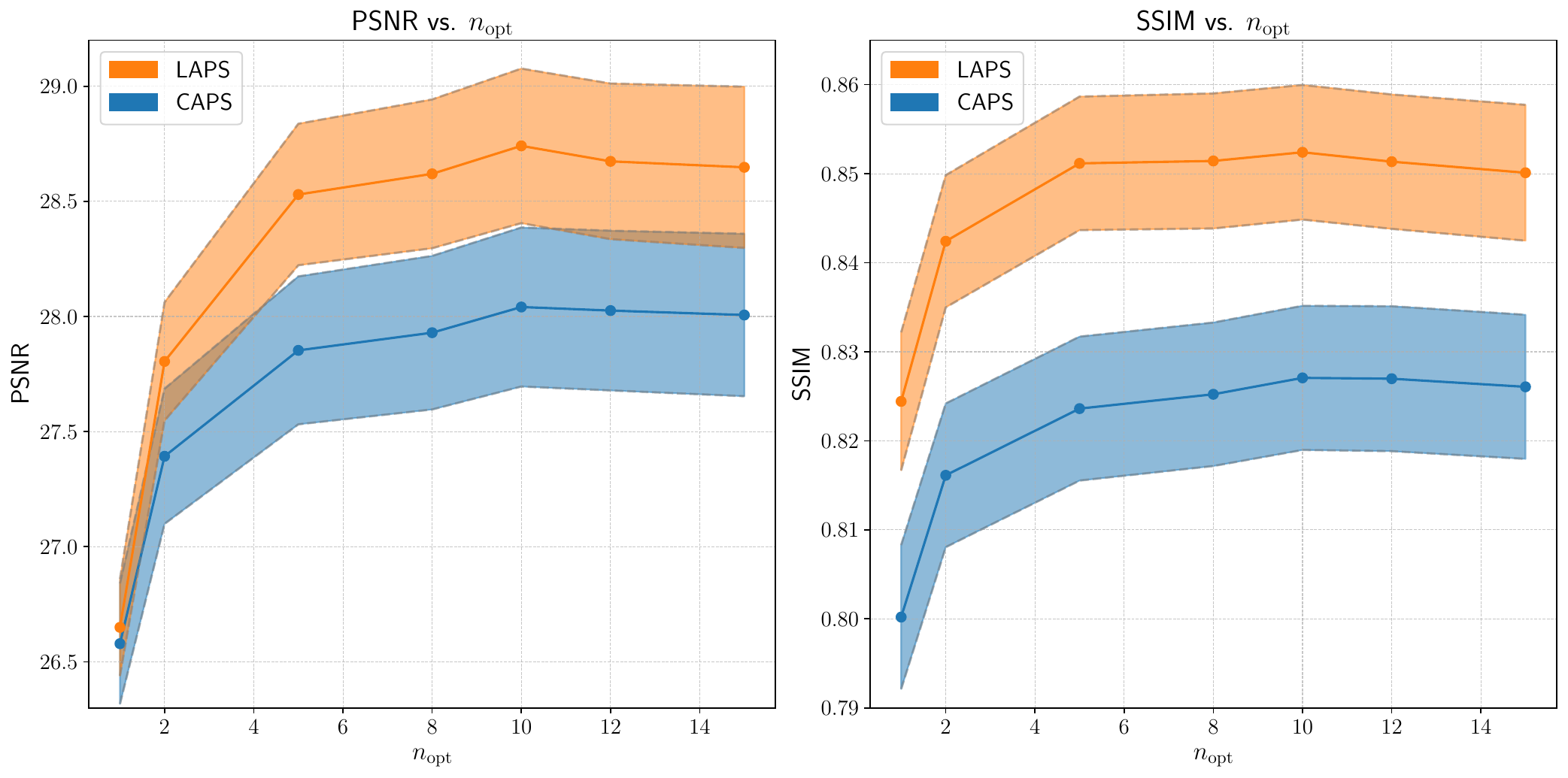}
    \caption{~Performance of reconstruction metrics for various $\nopt$, with $\tp$ chosen via $\myoperatorname{AutoInit}$, for $n_{\rm{step}} = 100$ DDIM steps. $\nopt=1$ equates to sampling with PLDS\cite{rout2023solving}; we observe that for both CAPS and LAPS, $\nopt=10$ maximizes PSNR and SSIM in the reconstruction test set. Additionally, LAPS consistently boosts performance compared to CAPS for each value of $\nopt$, showing the utility of the longitudinal prior.}
    \label{fig:nopt}
\end{figure}

\begin{figure*}[ht]
    \centering
    \includegraphics[width=0.8\linewidth]{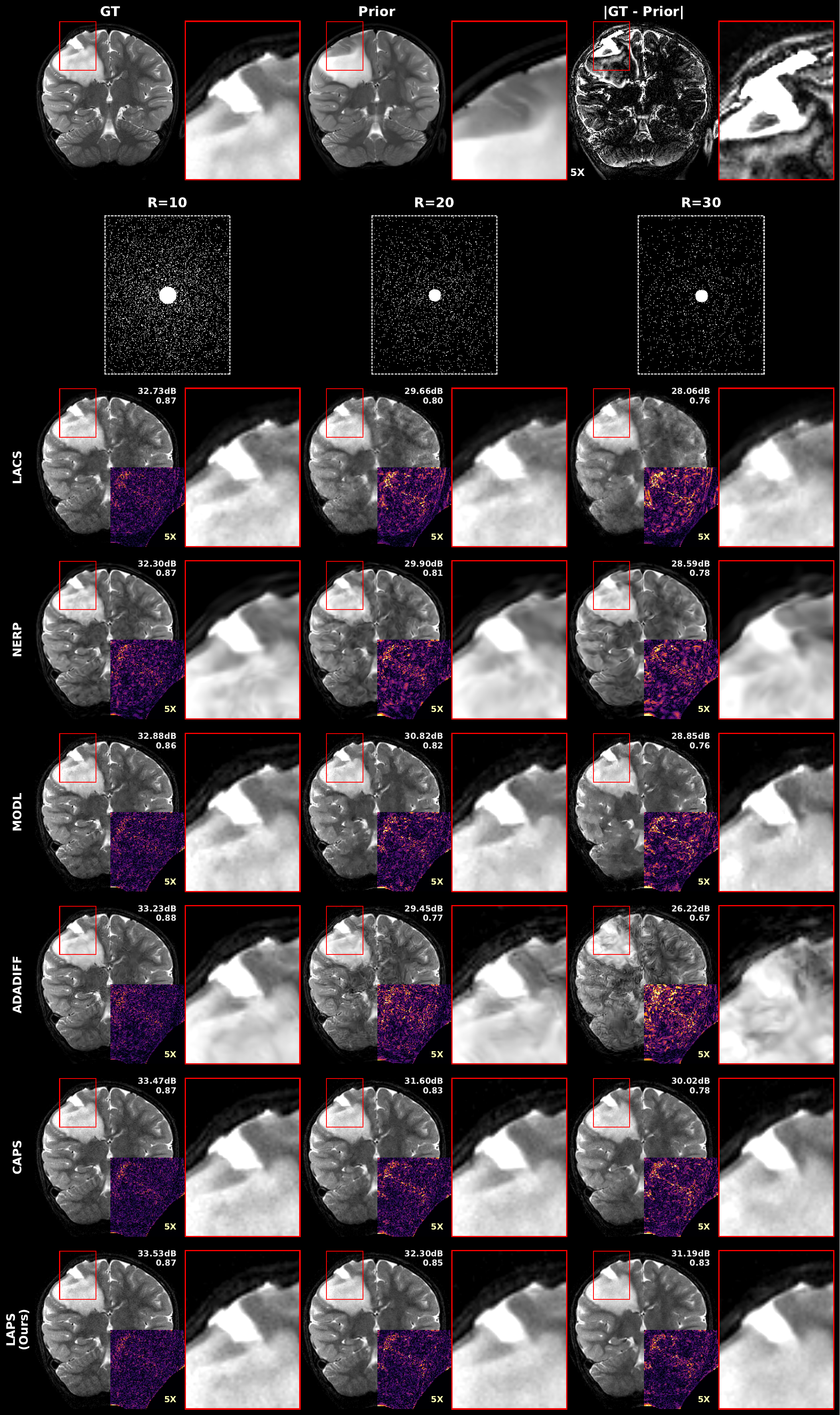}
    \vspace{-1em}
    \caption{~Reconstructions as in Fig.~\ref{fig:recon_accel_normal}, here shown for all methods.}
    \label{fig:recon_accel_abnormal_full}
\end{figure*}

\begin{figure*}
  \centering
  \begin{subfigure}[t]{0.4\textwidth}
    \centering
    \includegraphics[width=\linewidth]{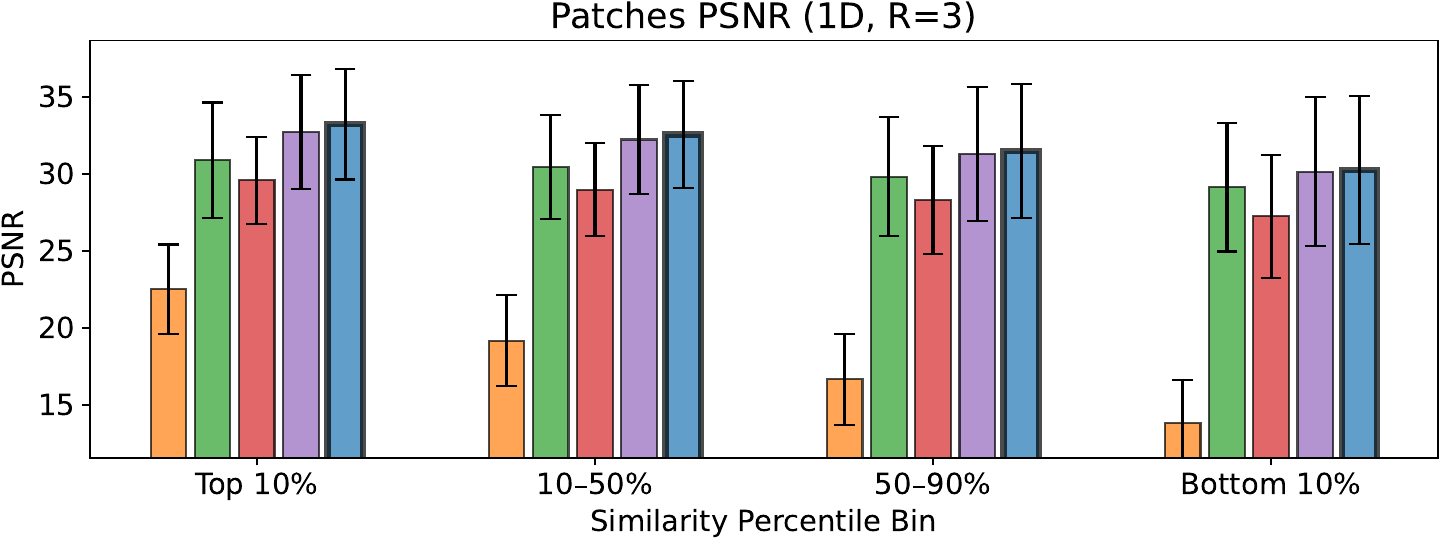}
    \caption{Patch-based PSNR histograms for 1D acceleration at $R=3$.}
    \label{fig:patch_metrics_sup_r3_psnr}
  \end{subfigure}%
  \hspace{1em}
  \begin{subfigure}[t]{0.4\textwidth}
    \centering
    \includegraphics[width=\linewidth]{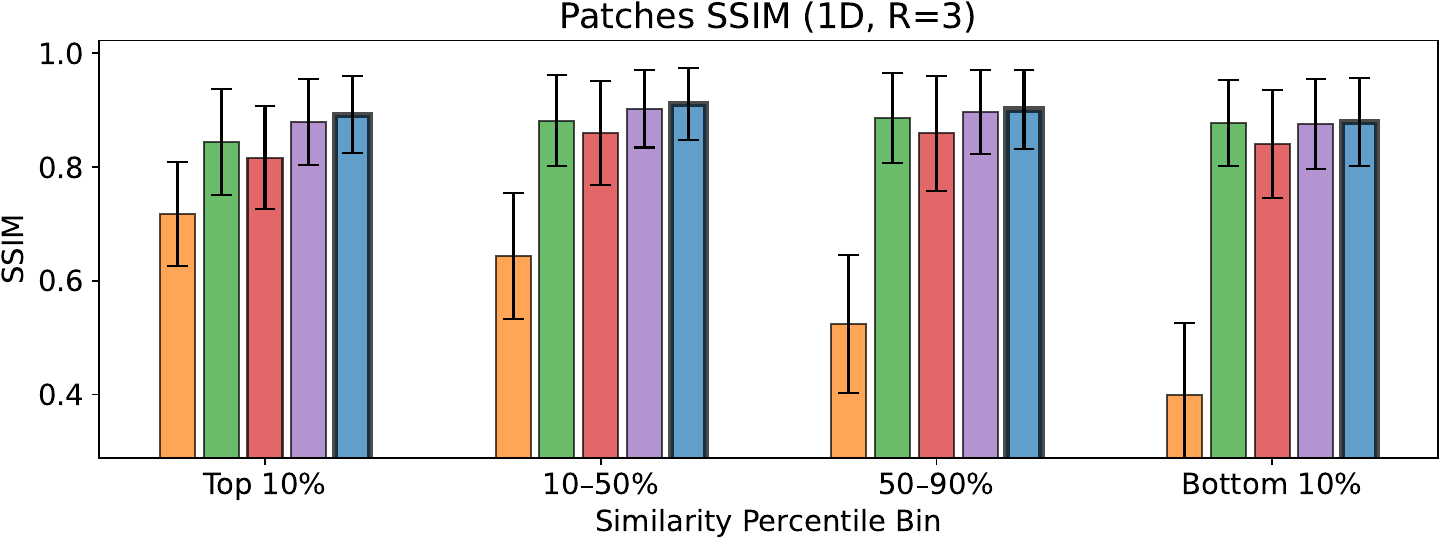}
    \caption{Patch-based SSIM histograms for 1D acceleration at $R=3$.}
    \label{fig:patch_metrics_sup_r3_ssim}
  \end{subfigure}
  \vspace{1ex}
  \begin{subfigure}[t]{0.4\textwidth}
    \centering
    \includegraphics[width=\linewidth]{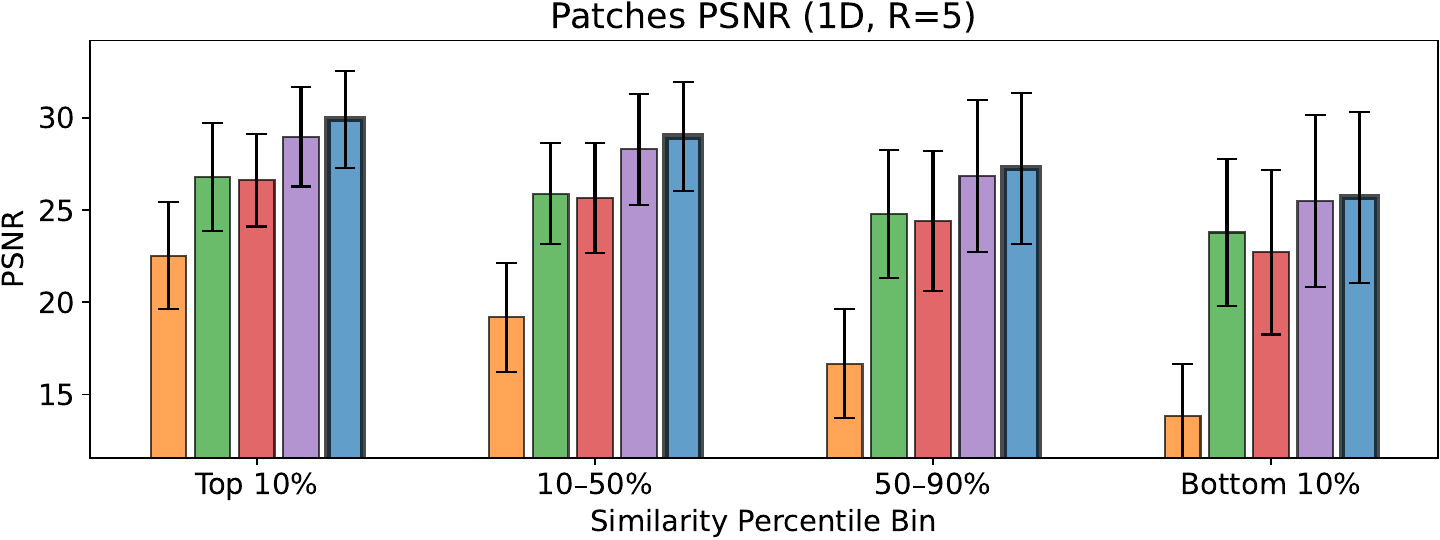}
    \caption{Patch-based PSNR histograms for 1D acceleration at $R=5$.}
    \label{fig:patch_metrics_sup_r5_psnr}
  \end{subfigure}%
  \hspace{1em}
  \begin{subfigure}[t]{0.4\textwidth}
    \centering
    \includegraphics[width=\linewidth]{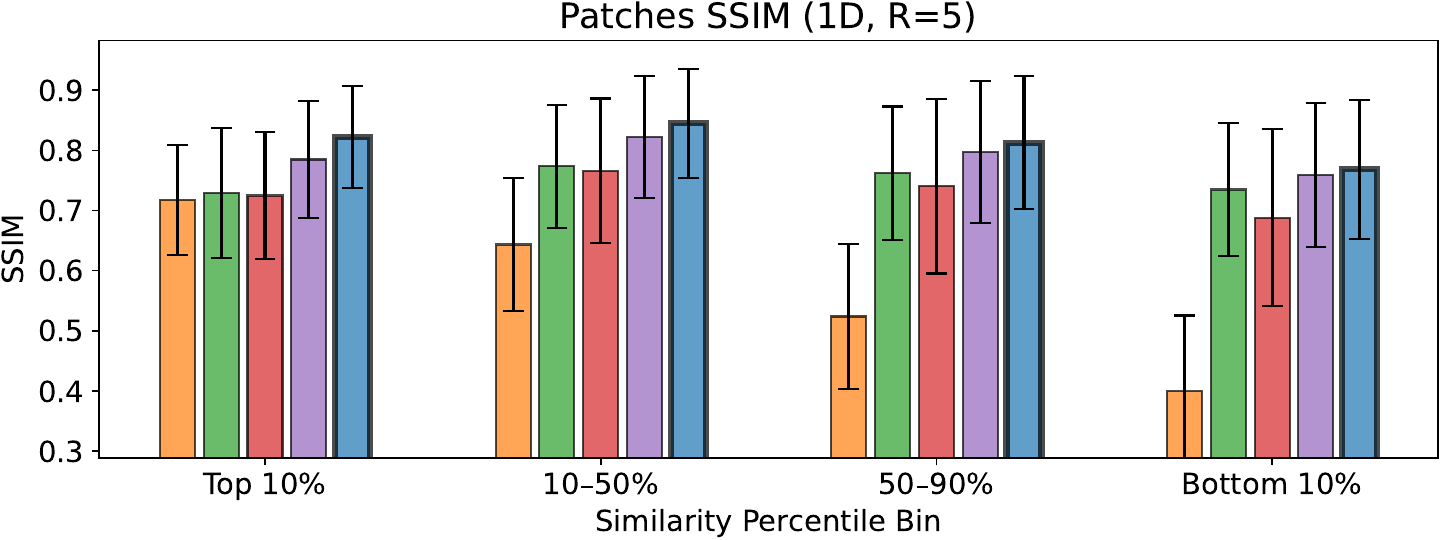}
    \caption{Patch-based SSIM histograms for 1D acceleration at $R=5$.}
    \label{fig:patch_metrics_sup_r5_ssim}
  \end{subfigure}
  \vspace{1ex}
  \begin{subfigure}[t]{0.4\textwidth}
    \centering
    \includegraphics[width=\linewidth]{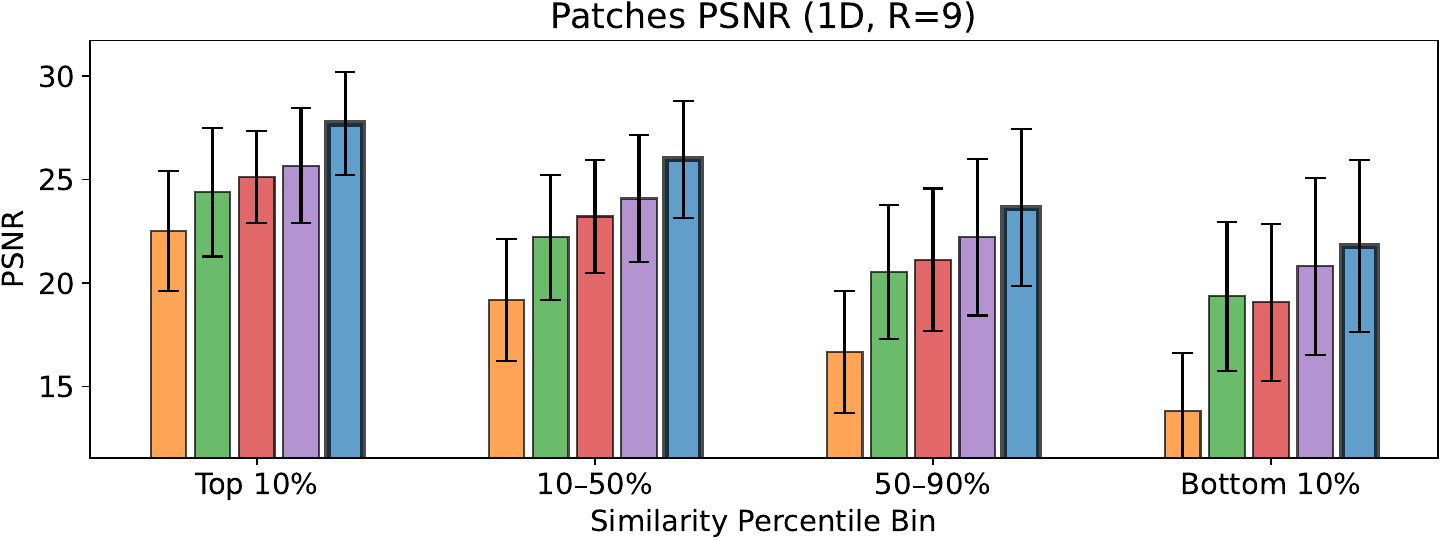}
    \caption{Patch-based PSNR histograms for 1D acceleration at $R=9$.}
    \label{fig:patch_metrics_sup_r9_psnr}
  \end{subfigure}%
  \hspace{1em}
  \begin{subfigure}[t]{0.4\textwidth}
    \centering
    \includegraphics[width=\linewidth]{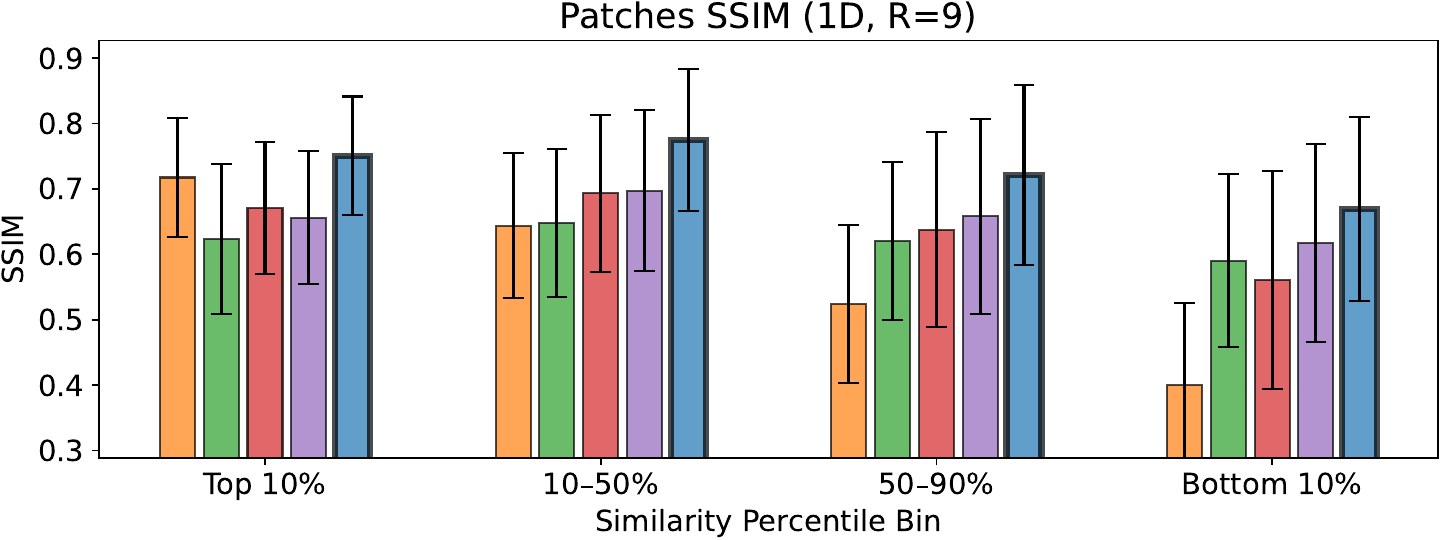}
    \caption{Patch-based SSIM histograms for 1D acceleration at $R=9$.}
    \label{fig:patch_metrics_sup_r9_ssim}
  \end{subfigure}
  \vspace{1ex}
\begin{subfigure}[t]{0.4\textwidth}
    \centering
    \includegraphics[width=\linewidth]{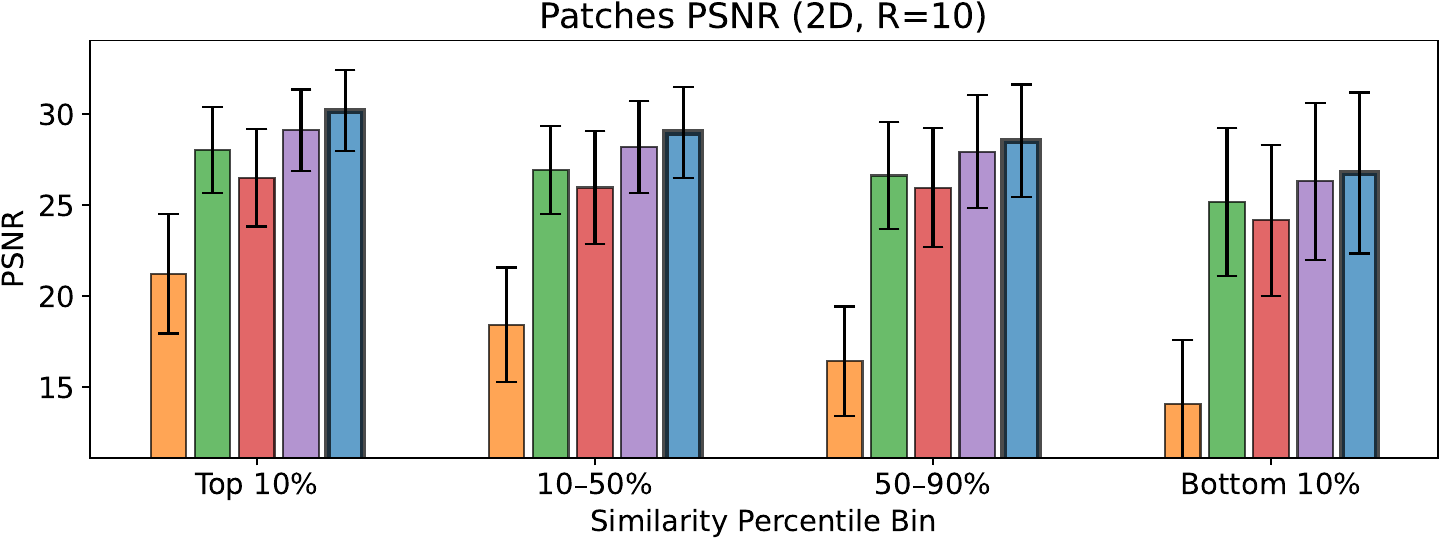}
    \caption{Patch-based PSNR histograms for 2D acceleration at $R=10$.}
    \label{fig:patch_metrics_sup_r10_psnr}
  \end{subfigure}%
  \hspace{1em}
  \begin{subfigure}[t]{0.4\textwidth}
    \centering
    \includegraphics[width=\linewidth]{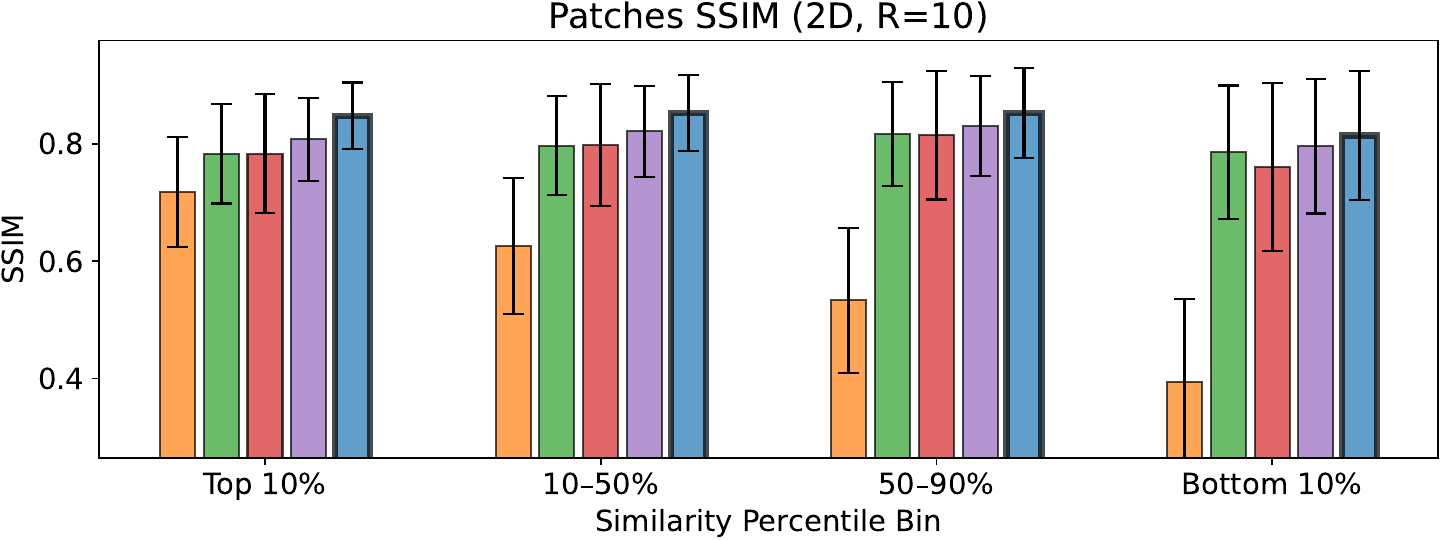}
    \caption{Patch-based SSIM histograms for 2D acceleration at $R=10$.}
    \label{fig:patch_metrics_sup_r10_ssim}
  \end{subfigure}
  \vspace{1ex}
  \begin{subfigure}[t]{0.4\textwidth}
    \centering
    \includegraphics[width=\linewidth]{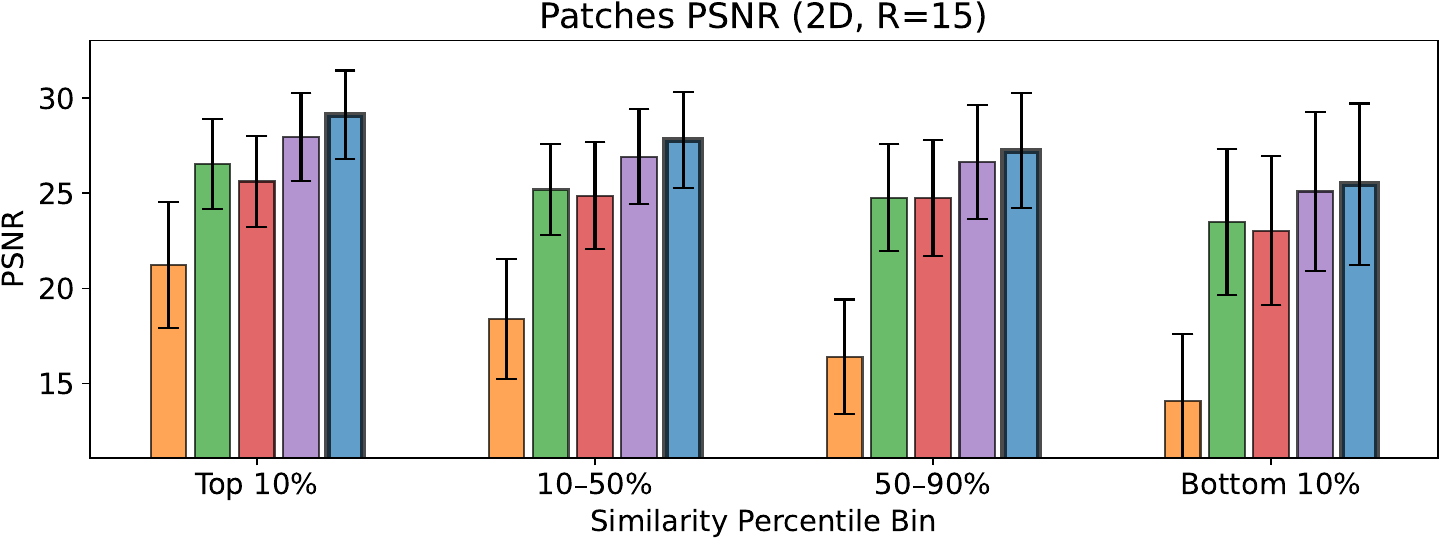}
    \caption{Patch-based PSNR histograms for 2D acceleration at $R=15$.}
    \label{fig:patch_metrics_sup_r15_psnr}
  \end{subfigure}%
  \hspace{1em}
  \begin{subfigure}[t]{0.4\textwidth}
    \centering
    \includegraphics[width=\linewidth]{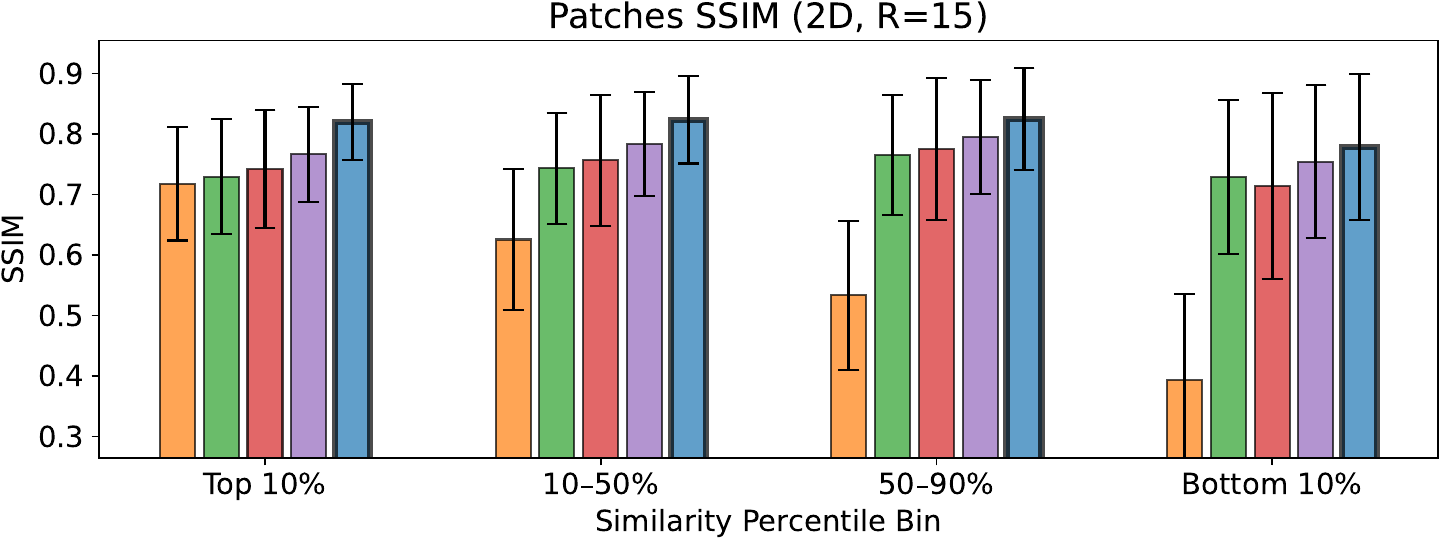}
    \caption{Patch-based SSIM histograms for 2D acceleration at $R=15$.}
    \label{fig:patch_metrics_sup_r15_ssim}
  \end{subfigure}
  \vspace{1ex}
  \begin{subfigure}[t]{0.4\textwidth}
    \centering
    \includegraphics[width=\linewidth]{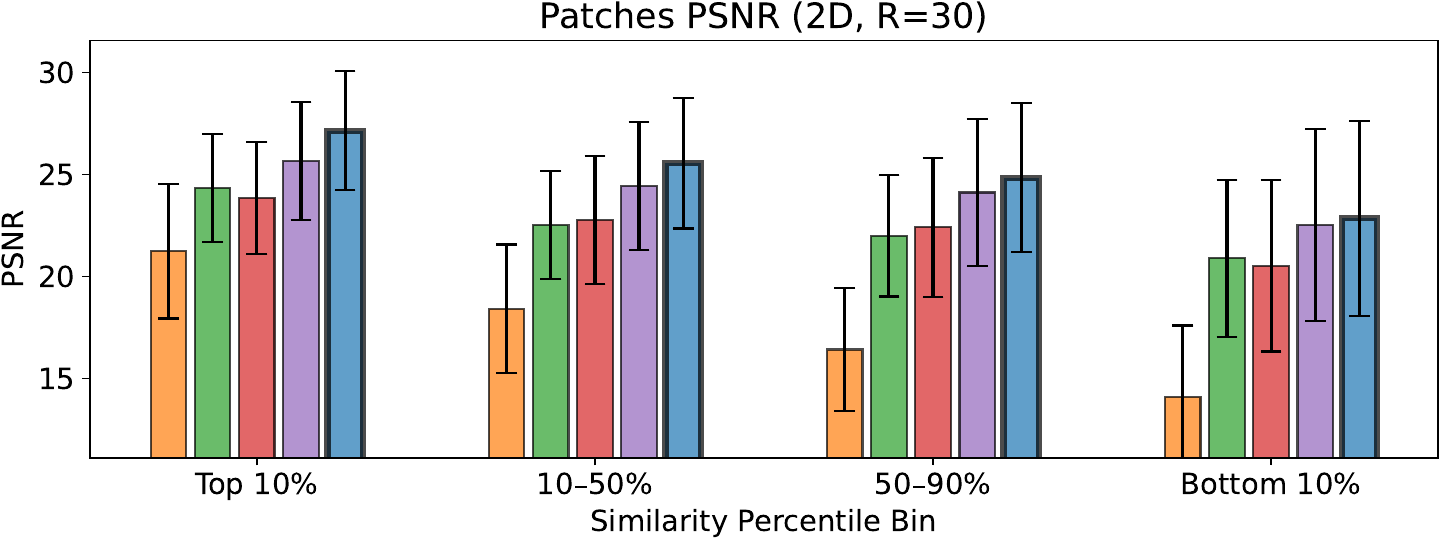}
    \caption{Patch-based PSNR histograms for 2D acceleration at $R=30$.}
    \label{fig:patch_metrics_sup_r30_psnr}
  \end{subfigure}%
  \hspace{1em}
  \begin{subfigure}[t]{0.4\textwidth}
    \centering
    \includegraphics[width=\linewidth]{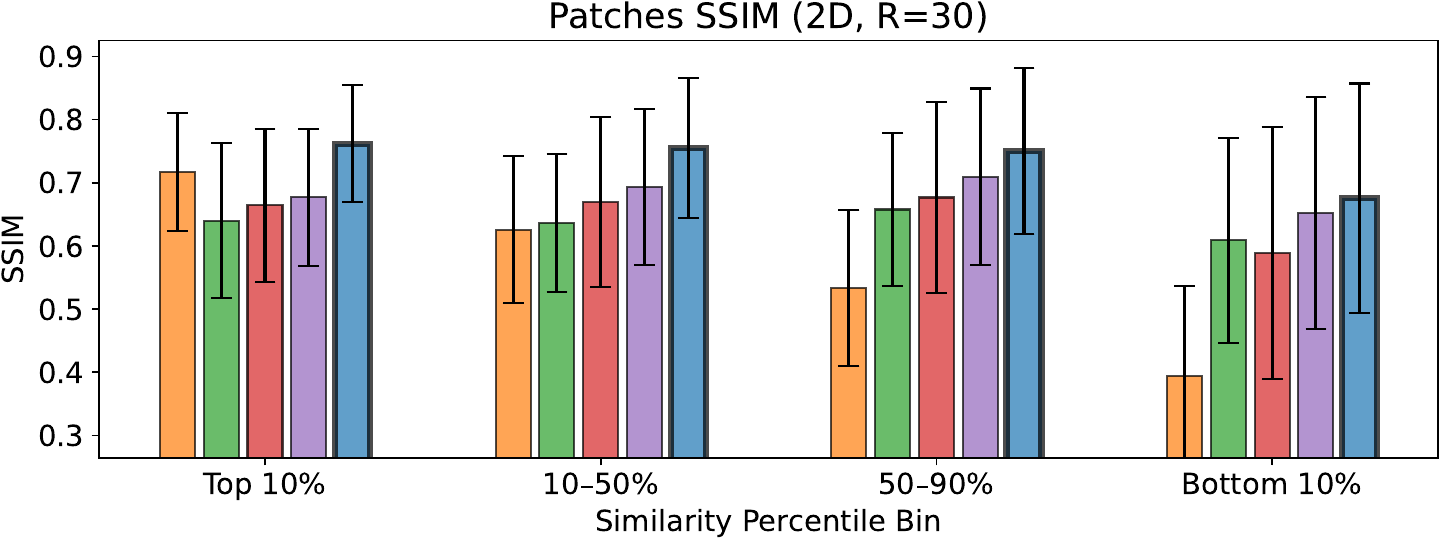}
    \caption{Patch-based SSIM histograms for 2D acceleration at $R=30$.}
    \label{fig:patch_metrics_sup_r30_ssim}
  \end{subfigure}
  \vspace{1ex}  
  \centering
  \begin{subfigure}[t]{0.9\textwidth}
    \centering
    \includegraphics[width=\linewidth]{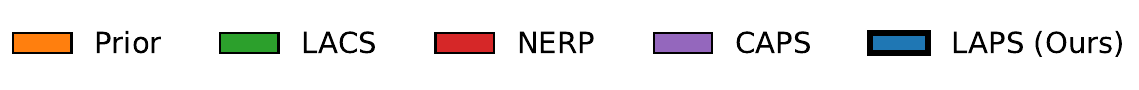}
    \label{fig:sub5}
  \end{subfigure}

  \caption{~Patch based histogram similar to the ones shown in Fig.~\ref{fig:patch_metrics}, for $R=3,5$ and $9$ (1D), and $R=10, 15$ and $30$ (2D).}
  \label{fig:patch_metrics_sup}
\end{figure*}

\begin{figure*}
    \centering
    \includegraphics[width=\linewidth]{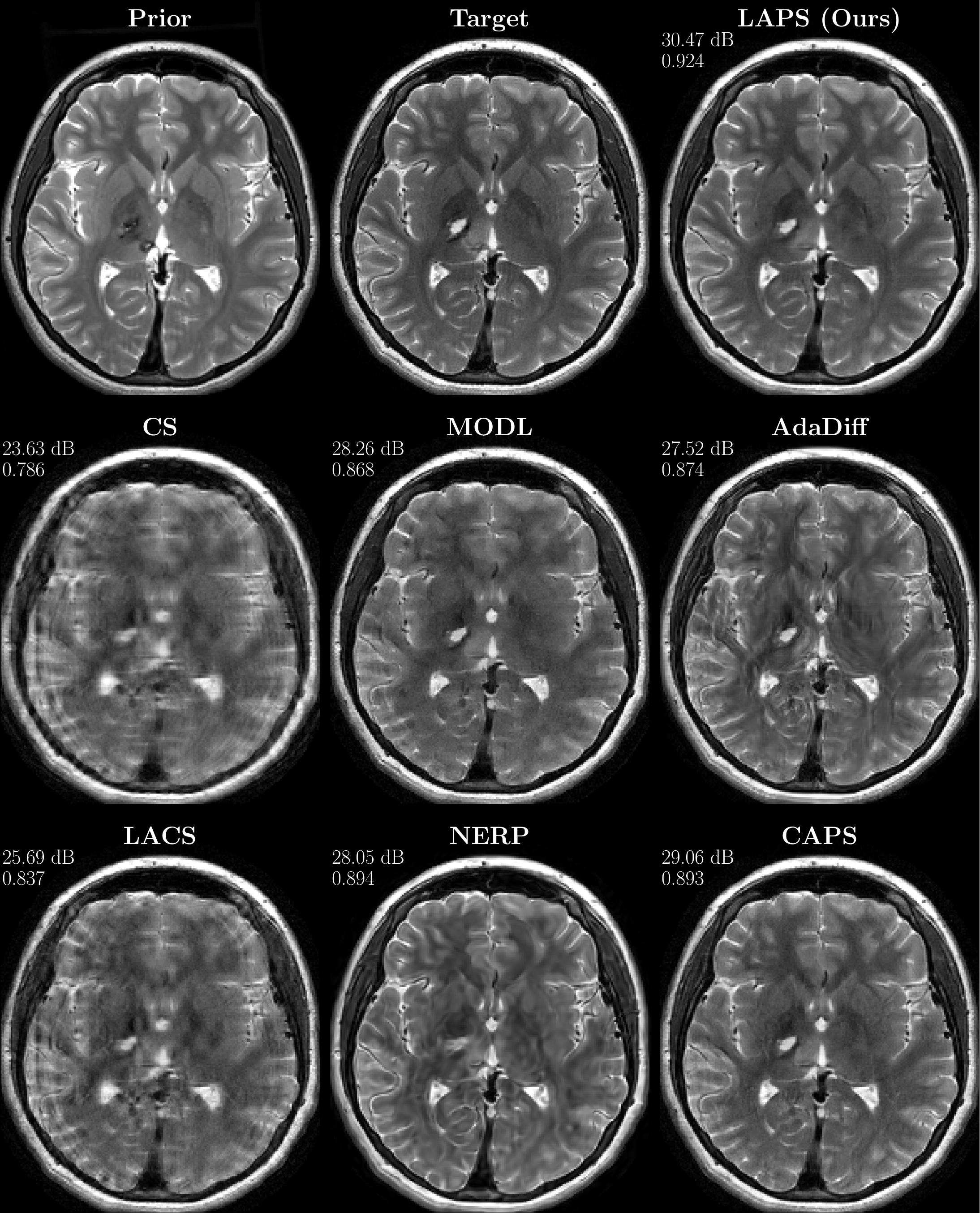}
    \caption{~Additional reconstruction example for all methods at R=7 with 1D undersampling, with PSNR and SSIM shown to the left of each image, shown as an axial view for the large change coronal example in Fig.~\ref{fig:diff_abnormal}.}
    \label{fig:highres_supp}
\end{figure*}

\begin{figure*}
    \centering
    \includegraphics[width=\linewidth]{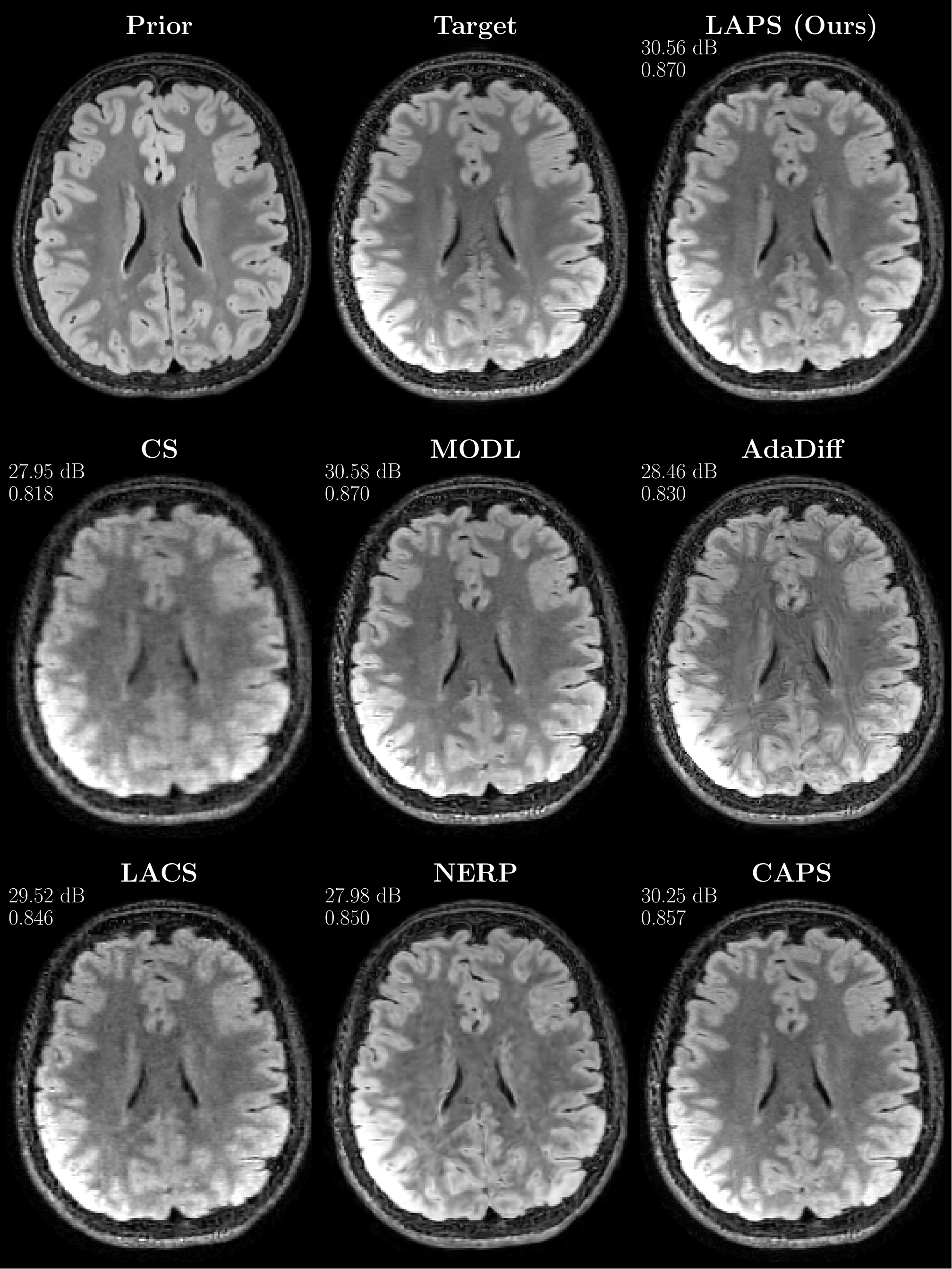}
    \caption{~Additional reconstruction example for all methods at R=20 with 2D undersampling, with PSNR and SSIM shown to the left of each image, for a $T_2$-FLAIR contrast.}
    \label{fig:highres_supp_t2f}
\end{figure*}

\begin{figure*}
    \centering
    \includegraphics[width=\linewidth]{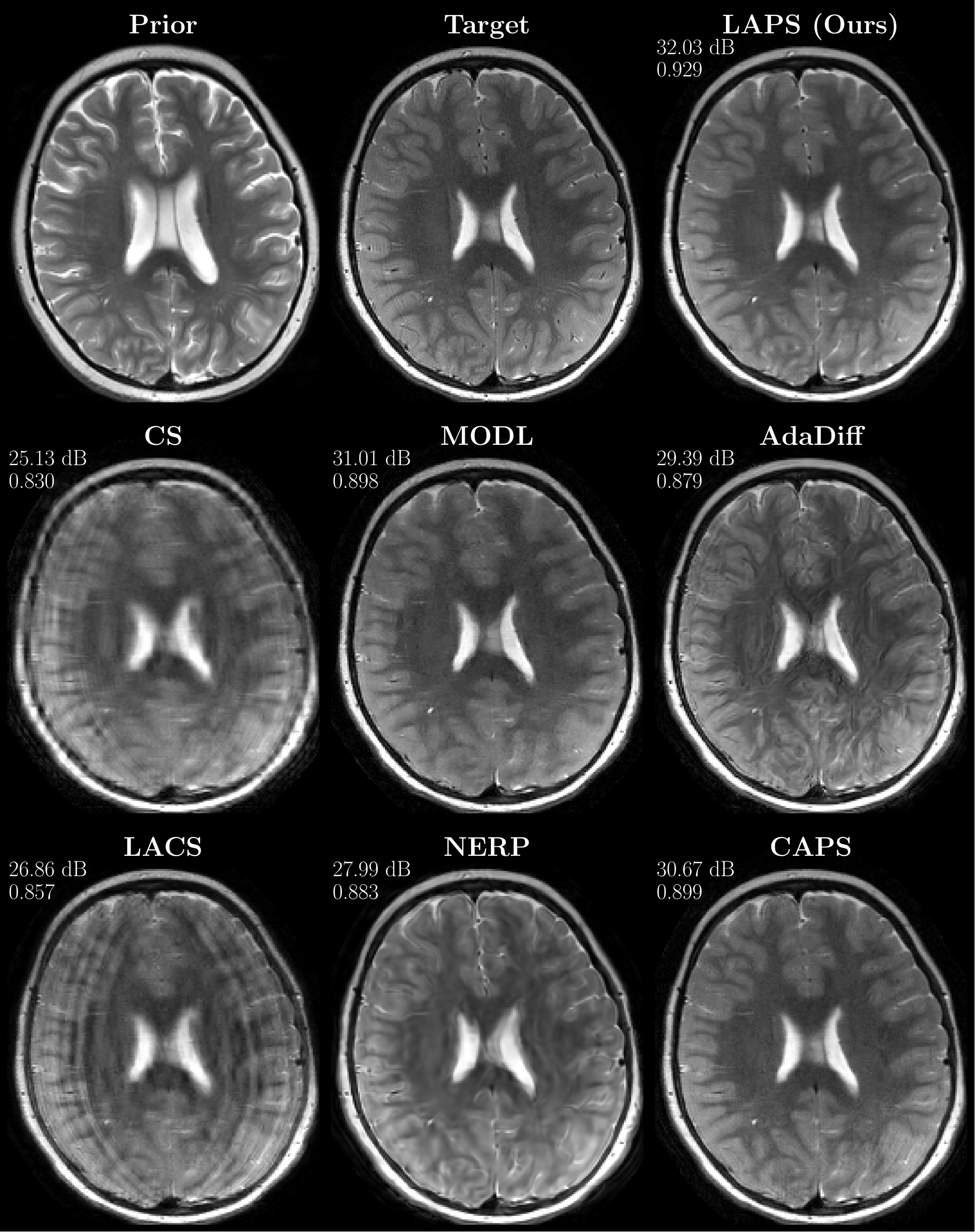}
    \caption{~Additional reconstruction example for all methods at R=7 with 1D undersampling, with PSNR and SSIM shown to the left of each image, where the prior contrast slightly differs from the new scan contrast.}
    \label{fig:highres_supp_contrast}
\end{figure*}

\end{document}